\documentclass[twocolumn,preprintnumbers]{aastex63} 
\usepackage{afterpage}
\usepackage{capt-of}
\usepackage{diagbox}

\usepackage[figuresright]{rotating}
\usepackage{amsmath}
\usepackage{amssymb,amsmath,latexsym,graphics, graphicx,epsfig,multirow,comment,hyperref,appendix,feyn,slashed,xcolor,afterpage, makecell} 
\usepackage{booktabs}
\usepackage{tabularx}
\usepackage{wasysym}
\usepackage{placeins}

\newcommand {\be} {\begin {equation}}
\newcommand {\ee} {\end {equation}} 

\newcommand {\bes} {\begin {equation*}}
\newcommand {\ees} {\end {equation*}}

\newcolumntype{L}[1]{>{\raggedright\let\newline\\\arraybackslash\hspace{0pt}}m{#1}}
\newcolumntype{C}[1]{>{\centering\let\newline\\\arraybackslash\hspace{0pt}}m{#1}}
\newcolumntype{R}[1]{>{\raggedleft\let\newline\\\arraybackslash\hspace{0pt}}m{#1}}

\usepackage{csvsimple}

\newcommand{\Z}{\mathbb{Z}}
\newcommand{\N}{\mathbb{N}}
\newcommand{\R}{\mathbb{R}}
\newcommand{\C}{\mathbb{C}}

\makeatletter
\newcommand\footnoteref[1]{\protected@xdef\@thefnmark{\ref{#1}}\@footnotemark}
\makeatother

\DeclareRobustCommand{\Sec}[1]{Sec.~\ref{#1}}

\DeclareRobustCommand{\App}[1]{App.~\ref{#1}}
\DeclareRobustCommand{\Tab}[1]{Table~\ref{#1}}

\DeclareRobustCommand{\Fig}[1]{Fig.~\ref{#1}}

\DeclareRobustCommand{\Eq}[1]{Eq.~(\ref{#1})}

\renewcommand{\be}{\begin{equation}}
\renewcommand{\ee}{\end{equation}}

\newcommand{\V}[1]{{ \bf #1}}

\newcommand{\Gaia}{\textit{Gaia}~}
\newcommand{\vesc}{v_{\rm esc}}
\newcommand{\vmin}{v_{\rm min}}

\newcommand{\vobs}{v_{\rm obs}}
\newcommand{\sigmaOut}{\sigma_{\rm{out}}}
\newcommand{\speed}{|\vec{v}|}

\definecolor{SeaGreen}{rgb}{0.13,0.55,0.33}

\definecolor{Magenta}{rgb}{0.75,0.15,0.33}

\begin{document}

\title{
Substructure at High Speed I: Inferring the Escape Velocity in the Presence of Kinematic Substructure
}

\author{Lina Necib}
\affil{Walter Burke Institute for Theoretical Physics,
California Institute of Technology, Pasadena, CA 91125, USA}
\affil{Center for Cosmology, Department of Physics and Astronomy,
University of California, Irvine, CA 92697, USA}
\affil{Observatories of the Carnegie Institution for Science, 813 Santa Barbara St., Pasadena, CA 91101, USA}

\author{Tongyan Lin}
\affil{Department of Physics, University of California San Diego, La Jolla, CA 92093, USA}

\begin{abstract}
The local escape velocity provides valuable inputs to the mass profile of the Galaxy, and requires understanding the tail of the stellar speed distribution. 
Following \cite{1990ApJ}, various works have since modeled the tail of the stellar speed distribution as $\propto (\vesc - v)^k$, where $\vesc$ is the escape velocity, and $k$ is the slope of the distribution. 
In such studies, however, these two parameters were found to be largely degenerate and often a narrow prior is imposed on $k$ in order to constrain $\vesc$. 
Furthermore, the validity of the power law form is likely to break down in the presence of multiple kinematic substructures.
In this paper, we introduce a strategy that for the first time takes into account the presence of kinematic substructure. 
We model the tail of the velocity distribution as a sum of multiple power laws without imposing strong priors.
Using mock data, we show the robustness of this method in the presence of kinematic structure that is similar to the recently-discovered \Gaia Sausage. 
In a companion paper, we present the new measurement of the escape velocity and subsequently the mass of the Milky Way using \Gaia DR2 data. \\
\end{abstract}

\section{Introduction} 
\label{sec:intro}

Evidence of the theory of hierarchical galaxy formation \citep{White:1977jf} has been abundant in recent years. The \Gaia mission \citep{2016A&A...595A...4L,2018arXiv180409365G,2020arXiv201201533G} in particular has found evidence of multiple mergers in the Milky Way (see e.g. \cite{2020ARA&A..58..205H} for a review). The Milky Way, and in particular the stellar halo, is a graveyard of disrupted substructure such as streams \citep{1999Natur.402...53H,Belokurov:2006ms,2007ApJ...658..337B,2020ApJ...901...48N}, clumps \citep{Diemand:2008in,2018MNRAS.475.1537M}, tidally disrupted dwarf galaxies (e.g. \cite{2006ApJ...647L.111B,2006ApJ...650L..41Z,2009MNRAS.398.1771N,2013ApJ...770...16K,2017MNRAS.467..573C,2017ApJ...838...11S}), and debris flow \citep{2018MNRAS.478..611B,2018Natur.563...85H,2018ApJ...863L..28M,2018ApJ...862L...1D,2019ApJ...874....3N,2018arXiv180704290L}.  
The \Gaia mission \citep{2016A&A...595A...4L,2018arXiv180409365G} has shed light on some of these substructures, and in particular led to the identification of a large debris flow called the \Gaia Sausage\footnote{In the remainder of this paper, we will refer to this substructure as the Sausage.} \citep{2018MNRAS.478..611B}, or \Gaia Enceladus \citep{2018Natur.563...85H}. Such a structure extends to $\sim 30$ kpc, including stars on highly eccentric orbits. It is most likely the product of a merging satellite of a stellar mass $10^{8-9} M_{\odot}$ that was disrupted at about redshift $z\sim 1-3$ \citep{2018ApJ...863L..28M,2018ApJ...862L...1D,2018arXiv180704290L}. 

In light of these findings, we must revisit the methods built for investigating properties of the Milky Way, and specifically in inferring the local escape velocity. Determining the escape velocity is important as it is used to: (1) constrain the total mass of the Milky Way, (2) predict signals for dark matter (DM) direct detection, (3) and build the DM density profile of the Milky Way. For instance, measurements of the escape velocity and the circular velocity can be used to constrain the potential of the Milky Way, assuming some spatial distribution of the disk, bulge, and dark matter. This has been done extensively in the literature (e.g. \cite{2007MNRAS.379..755S,2014A&A...562A..91P,2017MNRAS.468.2359W,2018A&A...616L...9M,2019arXiv190102016D}) in order to obtain a measurement of the Milky Way mass. However, the presence of many velocity substructures can affect our measurement of the escape velocity and thus the Milky Way mass. In this work, we aim to build a robust strategy for determining the escape velocity accounting for such substructure. In particular, we build a pipeline that incorporates, for the first time, multiple substructure components in modeling the tail of the stellar velocity distribution.

The majority of previous studies of the escape velocity are based on \cite{1990ApJ}, which model the tail of the stellar velocity distribution as 
\begin{align} \label{eq:fvtail} 
	f(v| \vesc, k ) \propto(\vesc - v)^k  \Theta(\vesc - v) \qquad v > \vmin,
\end{align}
where $v$ is the speed in Galactocentric coordinates and the two fitting parameters are the escape velocity $\vesc$ and the slope of the distribution $k$. This model is applied to stars with speeds greater than the threshold $\vmin$, and it is assumed that the approximation holds for $\vmin$ well above the local rotation speed. 

Following \cite{1990ApJ}, studies have inferred the local escape velocity using line-of-sight velocities with RAdial Velocity Experiment (RAVE) \citep{2007MNRAS.379..755S,2014A&A...562A..91P} and Sloan Digital Sky Survey (SDSS) \citep{2017MNRAS.468.2359W}, and then using 3D velocities from \Gaia \citep{2018A&A...616L...9M,2019arXiv190102016D}.  In all of these studies, there is a large degeneracy between $\vesc$ and $k$, as will be discussed further below. The degeneracy leads to rather large error bars on $\vesc$ and subsequently large error bars on the estimated mass of the Milky Way.

In order to overcome these large error bars, many of these works argued for narrow priors on the slope $k$;
the arguments for small values of $k$ were violent relaxation or collisional relaxation, both leading to $k\leq 2$ \citep{1990ApJ}. Meanwhile, \cite{2014A&A...562A..91P} (and subsequently \cite{2018A&A...616L...9M}) used cosmological simulations based on the Aquarius suite \citep{2008MNRAS.391.1685S,2009MNRAS.396..696S} to argue for a prior $k \in [2.3, 3.7]$, while \cite{2019arXiv190102016D} used the Auriga simulation \citep{2017MNRAS.467..179G} to argue that for mergers resembling the Sausage, $k$ should be small, and therefore $k \in [1, 2.5]$.  More recently, \cite{2020arXiv200616283K} used a much larger sample of stars with only proper motion measurements to reduce the degeneracy. However, a difficulty in using only proper motions is that the tail of the distribution is not necessarily populated all the way up to $\vesc$, with \cite{2020arXiv200616283K} estimating a possible 10\% bias.

These studies illustrate some of the difficulties in using \Eq{eq:fvtail} to model the tail. The degeneracy in the parameters $\vesc$ and $k$ is due to the fact that a higher $\vesc$ can be partially compensated by a higher slope $k$ in the shape of the distribution. Because there are very few stars near $\vesc$, a fit to \Eq{eq:fvtail}  could then easily lead to biased results if the model is not a good description of the data over the entire range of speeds. This could be the result of additional unbound stars, a mismodeled or unmodeled component, or measurements with large errors that contaminate the data set.  A second related issue is that there is no precise definition of the ``tail'' where the model is expected to be a good description. For example, \cite{2019MNRAS.487L..72G} studied numerical simulations and found that those distributions deviate from \Eq{eq:fvtail} due to the presence of substructure, often leading to underestimates of the Milky Way mass.

In this paper, we argue for an approach that can more robustly determine where the ``tail'' of the stellar speed distribution is, and that takes into account the presence of kinematic substructure. Given what is known about the Sausage, it is likely that a large fraction of stars in the tail of the distribution can be attributed to this substructure; as argued by \cite{2019arXiv190102016D}, it will have a different slope $k$ than the rest of the stellar halo. Including substructure is thus physically motivated. The tail of the distribution is then the sum of (at least) two distributions and might not be well-described by a single power law for low $\vmin$. Not including substructure could then bias $\vesc$ measurements. 

To address these points, we build a pipeline where we add a second component of the velocity distribution, also modeled as in \Eq{eq:fvtail} but with a new slope $k_S$. While this is motivated by substructure, it can also be viewed as a more flexible model for the steeply falling speed distribution. We show how $\vesc$ can be obtained more robustly by performing tests on the data as a function of $\vmin$ and the number of bound components. For instance, it is expected that a single component will be adequate for large enough $\vmin$. Performing these tests can ensure that the fit for $\vesc$ is not biased by structure in the speed distribution at lower speeds. 
 In all of these tests, it is important that we keep the priors for all parameters as wide as possible so we are not artificially shaping the results.

In this work, we present the pipeline and analyses with mock data sets containing kinematic substructure. 
This paper is organized as follows. In \Sec{sec:motivation}, we first discuss in more depth the motivation for including substructure and illustrate the main points. The details of the pipeline are provided in \Sec{sec:analysis}. In \Sec{sec:simulations} we test the pipeline on mock data sets containing substructure, and compare results when one or two bound components are used in the fit.  We also study the effect of strong priors on the results, and test the robustness of the fits when the slopes in the components are changed. In a companion paper \citep{data_escape}, we apply this method on \Gaia DR2 data for stars in the local neighborhood ([7,9]~kpc in Galactocentric distance), to present the most robust estimate of the local escape velocity, from which we deduce the mass of the Milky Way.

\section{Motivation for including substructure \label{sec:motivation} }

In this section, we discuss two broad motivations for including kinematic substructures in modeling the tail of the velocity distribution. First, kinematic substructure ---the Sausage--- is present in the Milky Way \citep{2018MNRAS.478..611B,2018Natur.563...85H} and can comprise a large fraction of stars \citep{2019ApJ...874....3N}. However, based on empirical studies of the Sausage kinematic properties from \cite{2019ApJ...874....3N}, it is not obvious what the substructure slope $k$ and fractional contribution to the tail of the velocity distribution should be. Rather than using simulations as a prior on the slope, we prefer to obtain independent information about the kinematic substructure from the data. 

Second, not accounting for this substructure can lead to biases in Milky Way mass estimates \citep{2019MNRAS.487L..72G}.
In particular, the choice of a low $\vmin = 300$~km/s is common in the literature as it increases statistics. For such low $\vmin$ compared to an expected escape velocity $\vesc \sim 500$ km/s, there may be contributions from multiple kinematic structures, including for example the Sausage. Not accounting for the second component can then pull the fit towards larger $\vesc$, depending on $\vmin$. By performing a two-component fit over different $\vmin$, we can demonstrate the robustness of the posterior distributions on $\vesc$ and hence on Milky Way mass estimates. 

\subsection{The presence of substructure in the tail}

\begin{figure*}[t] 
   \centering
	\includegraphics[width=0.45\textwidth]{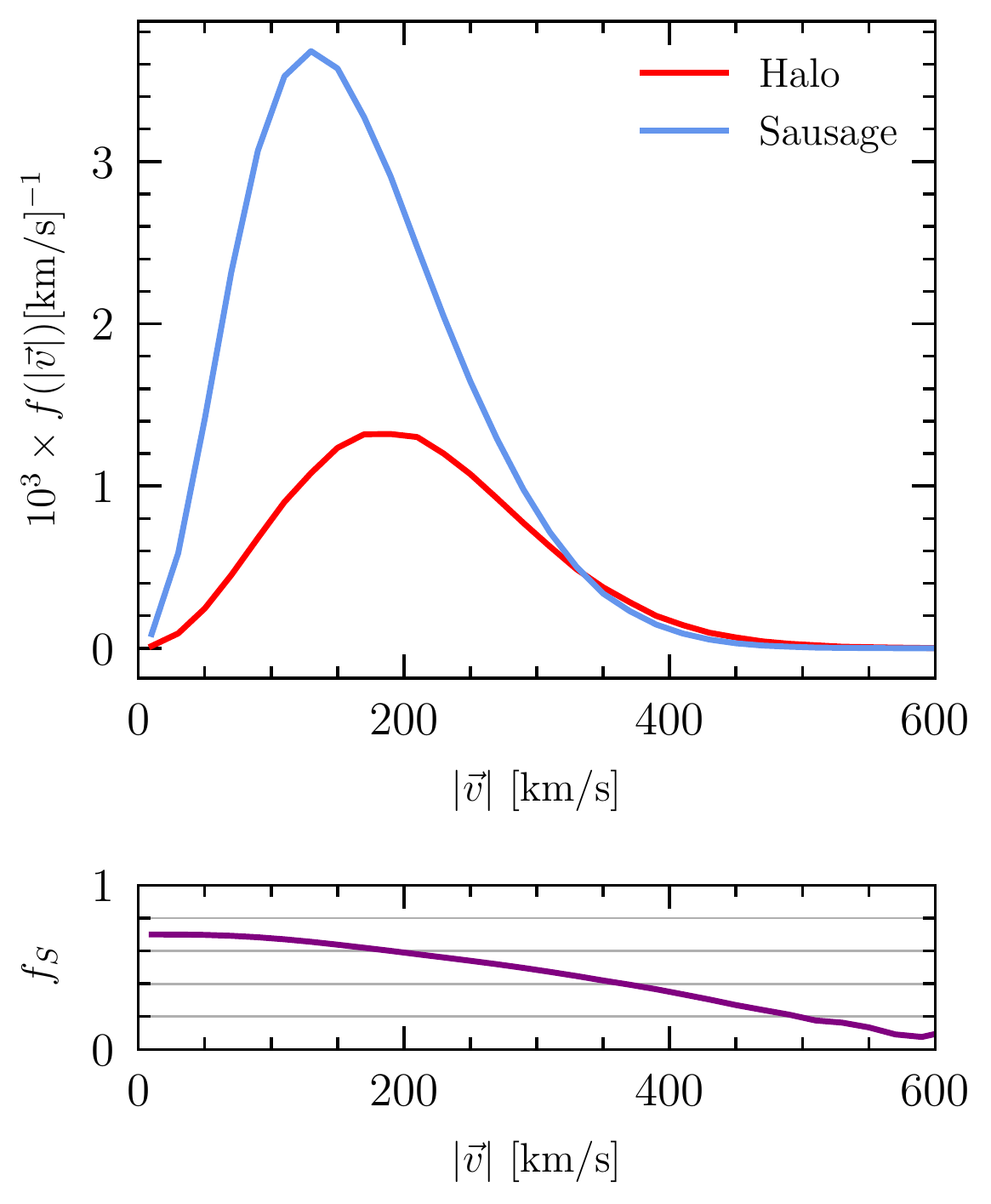} 
	\includegraphics[width=0.45\textwidth]{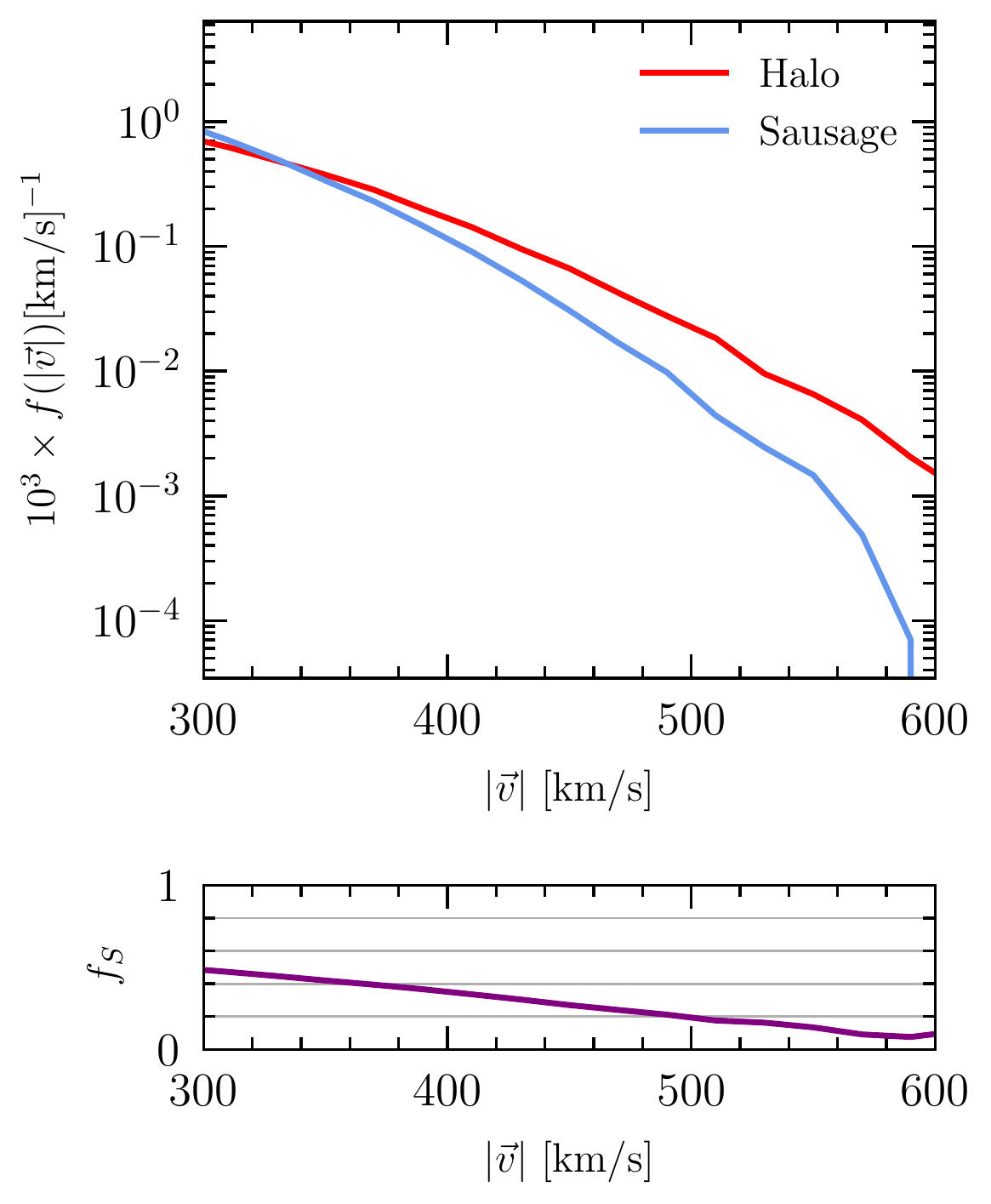} 
   \caption{Speed distributions of the Halo and Sausage components from \cite{2019ApJ...874....3N}, assuming a total Sausage fraction of 70\% of the sum of the distributions. (\textbf{Left}) Full speed distributions, (\textbf{Right}) speed distributions above $300$ km/s in logarithmic space. (\textbf{Bottom}) Fractional contribution of the Sausage distribution for all speeds above $\speed$ (see Eq.~\eqref{eq:fraction}).}
   \label{fig:distributions_SDSS}
\end{figure*}

The Milky Way recently underwent a major merger, the Sausage \citep{2018MNRAS.478..611B,2018Natur.563...85H}, which was discovered through its distinct chemical and phase space properties. It was shown that the merger contributes about $\sim 60-72\%$ of the non-disk stars in the local neighborhood \citep{2019ApJ...874....3N,2019arXiv190707681N}, which means that it would be expected to strongly shape the tail of the stellar speed distribution.

To illustrate this, in \Fig{fig:distributions_SDSS} we plot the speed distributions of the stellar Halo and the Sausage \citep{2019ApJ...874....3N}, normalized by their relative fractions, where we assume that the Sausage comprises $70\%$ of the total distribution.\footnote{\url{https://linoush.github.io/DM_Velocity_Distribution/}} These are the best fit distributions that have been built by modeling Galactocentric velocities and metallicity measurements from a cross match of \Gaia DR2 and the Sloan Digital Sky Survey \citep{2012ApJS..203...21A} using a Gaussian Mixture model with Halo, Sausage, and Disk components. While the Halo and Disk were modeled as three-dimensional Gaussian distributions in spherical Galactocentric coordinates, the Sausage was modeled as a sum of two Gaussians with opposite-sign means and equal dispersions in $v_r$, and single Gaussians in $v_{\theta}$ and $v_{\phi}$. 

In the left panel of \Fig{fig:distributions_SDSS}, we plot the full speed distributions, while on the right we plot the tail of the distributions on a log scale, with $\speed>300$ km/s. In the bottom panels, we plot the integrated ratio of the Sausage distribution to that of the sum of the Halo and Sausage distributions:
\be \label{eq:fraction}
f_S(\speed)= \frac{\int_{\speed}^{\infty} \mathbf{f_S}(v') dv'}{\int_{\speed}^{\infty} \mathbf{f_S}(v') dv' + \int_{\speed}^{\infty} \mathbf{f_H}(v') dv'},
\ee
where $\mathbf{f_S}$ and $ \mathbf{f_H}$ are the speed distributions of the Sausage and the Halo respectively. 
 We see that the fraction of the Sausage distribution decreases as a function of speed. This is because the distribution for the Sausage peaks at lower velocities than that of the Halo. From the right panel, we also see that the slope of the Sausage is larger\footnote{To give some intuition, $k \rightarrow 0$ is a Heaviside function that is truncated at $v = \vesc$, while $k \rightarrow \infty$ is a sharply falling function.} than that of the Halo ($k_S > k$), corresponding to a sharper drop at higher velocities.  Of course, the distributions in \Fig{fig:distributions_SDSS} are the result of a fit to a Gaussian Mixture Model, and therefore are not tuned for accuracy of the tails.  Nevertheless, based on this, one might still expect a sizable percentage of the tail of the distribution to be coming from the Sausage.

Because the contribution of the Sausage for a given $\vmin$ is {\emph{a priori}} unknown but likely sizable, single-component fits to the tail of the velocity distribution might fail to describe the data accurately and bias results. Based on studies of the Sausage so far, there is not enough information to set a strong prior on the slope or fractional contribution of the Sausage.   \cite{2019arXiv190102016D} adopts a prior on $k \in [1, 2.5]$ based on simulations where the tail of the velocity distribution is dominated by a substructure like the Sausage, which might not be true in the case of the Milky Way.  Instead, we find the opposite behavior in \Fig{fig:distributions_SDSS}. Meanwhile, other works have argued for and used different priors based on simulations, with \cite{2007MNRAS.379..755S} using $k \in [2.7, 4.7]$ and \cite{2014A&A...562A..91P} arguing that one expects $k \in [2.3, 3.7]$. The choice of prior thus largely depends on the merger history of the simulations considered. For a fit with degenerate parameters, the results are then strongly molded by the priors, and could lead to incorrect results, as we discuss in \Sec{sec:simulations}. In this work, we prefer to remain agnostic as to the interpretation of each bound distribution. Instead, we will show how including kinematic substructure in the fit allows an independent robust way to measure properties of the substructure.

\subsection{Influence of substructure on single-component fits to the tail}

\begin{figure*}[t] 
   \centering
	\includegraphics[width=0.48\textwidth]{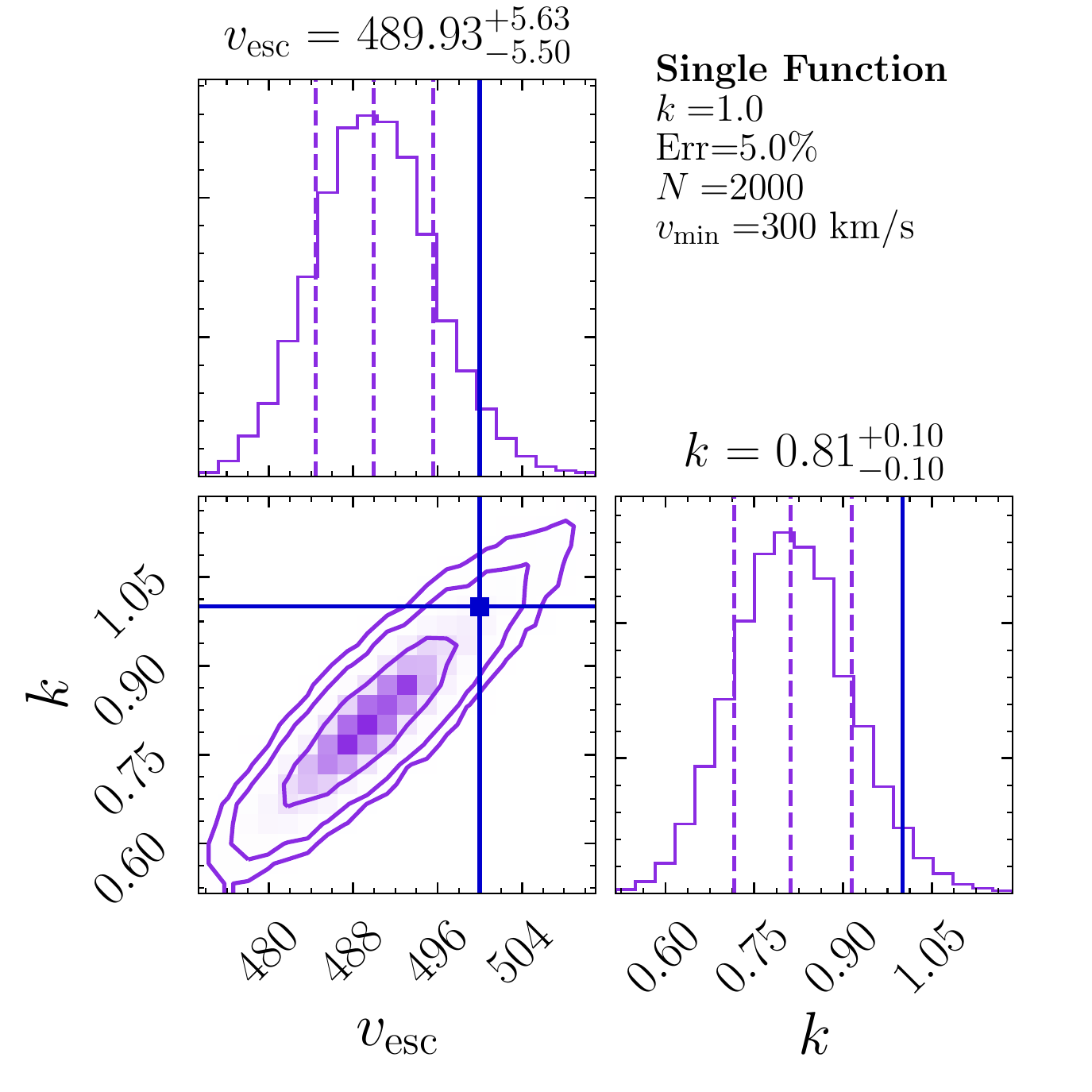} 
	\includegraphics[width=0.48\textwidth]{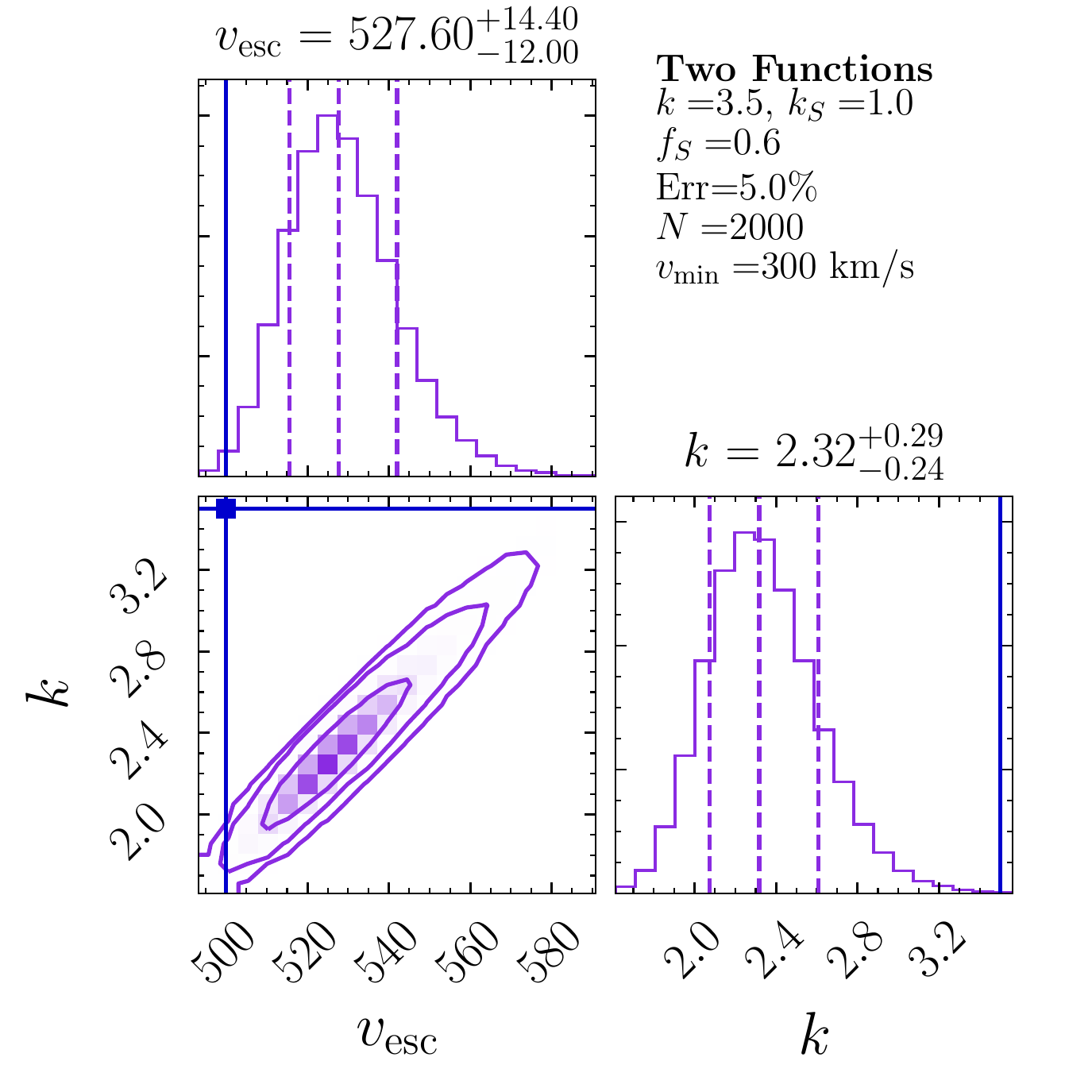} 
   \caption{Corner plots of the fit to a single bound component. The 2D contours are the 68\%, 95\%, and 99\% containment regions. The dashed lines in the 1D posteriors are the median and 1$\sigma$ containment regions. ({\bf Left}) We use mock data that includes a single component with true values $\vesc = 500$ km/s and $k = 1$, which are indicated by the blue lines. The stars were smeared with a random Gaussian error of 5\%. The fit is consistent with the true values. ({\bf Right}) We use mock data that contains two bound components, as described in more detail in the text. In this case, there is a larger degeneracy between $k$ and $\vesc$ compared to the single-component data and $\vesc$ is biased toward larger values.  }
   \label{fig:corner_plot}
\end{figure*}

Having argued that there could be a large substructure component, we now show how results could be affected if this substructure is not included in the model. While we defer the detailed discussion of our pipeline to \Sec{sec:analysis}, here we show the results of some analyses on mock data to illustrate the main ideas. 

We generate two sets of mock data, one with a single bound distribution drawn from \Eq{eq:fvtail} and another containing two  velocity distributions with different slopes $k$ but a common $\vesc$. For the single bound distribution, we assume $k=1$ and $\vesc = 500$ km/s. For the mock data set with substructure, we assume a substructure fraction of $f_{S}$ = 0.6 and substructure slope $k_{S} =1$, while the other component has $k=3.5$. In both cases, we also include an unbound outlier population, which is a fraction $f = 0.01$ of the total stars and described by a Gaussian with dispersion $\sigma_{\rm out} = 1000$ km/s. To simulate a realistic data set, we take $v_{\rm min} = 300$ km/s and smear the true speeds of the stars with a random Gaussian error of 5\%, which will be further discussed later.  

In \Fig{fig:corner_plot}, we show the result of fitting both these data sets to a model with a single bound component. The model also includes the outlier component and accounts for the errors on the velocity measurements, as will be described in the next section. When the mock data contains a single component (left), we find that the fit is indeed consistent with the true model. 

However, when the mock data contains two components (right panel of \Fig{fig:corner_plot}), we find two important impacts on the fit for $\vesc$: the best fit value of $\vesc$ is biased higher, and the degeneracy between $k$ and $\vesc$ is larger, with larger error bars on $\vesc$ and $k$. To understand this behavior, we first observe that the best-fit slope $k$ is in between the values of the individual slopes of the two components; the best fit value is $k = 2.32^{+0.29}_{-0.24}$ while the true values are $k_S = 1$ and $k = 3.5$.  This is because the tail of the distribution will be described by the component with lower slope (here $k=1$), while the stars at lower speeds will drive the preferred slope to higher values.  Because of the correlation between the effects of increasing $k$ and $\vesc$, the escape velocity will also be driven to larger values to better fit the tail.

As discussed in \Sec{sec:intro}, other works have set strong priors on the slope $k$ in order to reduce the degeneracy seen between $k$ and $\vesc$. In general, these priors can lead to a nonconvergent fit, with the slope parameter $k$ piling up at the edge of the priors.  Choosing a narrow prior thus shapes the posteriors, leading potentially to incorrect results. Indeed, \cite{2019arXiv190102016D} found that the choice of the priors affects the end result of the Milky Way mass. The example of \Fig{fig:corner_plot} shows that the degeneracy in $k$ and $\vesc$ could be partially due to the presence of multiple stellar components. As we will show in detail throughout this work, modeling both components properly leads again to a robust fit.  The goal of this paper is thus to build a robust method that accounts for kinematic substructure, and is less sensitive to the choice of priors.

\section{Analysis}
\label{sec:analysis}

\subsection{Multi-Component Pipeline}
\label{sec:standard}
We now present the likelihood function for an unbinned analysis on a sample of stars with minimum observed velocity $\vmin$.  This analysis will involve either a single  bound component, or two bound components. 
Each stellar velocity distribution will be modeled as in \Eq{eq:fvtail}, above a minimum velocity threshold $\vmin$. 

The true velocity distribution is smeared out by the measurement error, modeled by a 1-dimensional Gaussian. We define for each star $\alpha$ the probability to observe it at velocity $\vobs$
\be
	 p_\alpha(\vobs|v) = \frac{1}{\sqrt{2 \pi \sigma_{v,\alpha}^2}} \exp \left( - \frac{(v   - \vobs)^2}{2 \sigma_{v,\alpha}^2} \right),
	 \label{eq:pvvobs}
\ee
where $v$ is the true velocity of the star, and $\sigma_{v,\alpha}$ is the observed measurement error. Then the likelihood for star $\alpha$ to be drawn from the distribution defined by \Eq{eq:fvtail} is given by
\begin{align}  \label{eq:tail_error}
	\tilde p_\alpha(\vobs | \vesc, k) = & \, C_\alpha(k, \vesc)  \times \nonumber \\
	   \int_{0}^{\infty}  \, dv & \, (\vesc - v)^k \,  p_\alpha(\vobs|v) \Theta(\vesc - v)
\end{align}
where the lower limit of the integration region is 0 to account for stars with \textit{true} speed below $\vmin$. The factor $C_\alpha(k, \vesc)$ leads to a normalized PDF in the {\emph{data}} region $[\vmin, \infty]$, meaning
\begin{align}
	\int_{\vmin}^{\infty} d \vobs \, \tilde p_\alpha(\vobs | \vesc, k) = 1.
\end{align}
Note that the power law in \Eq{eq:fvtail} should approximately describe stars even with velocities below $\vmin$, since some of these stars may be observed above $\vmin$ due to measurement error. 
The normalization factor is 
\begin{align}
	   C_\alpha(k, \vesc) =  &2 \left[  \int_{0}^{\vesc} \ dv (\vesc - v)^k  \times \right. \nonumber \\
	  & \left.  \left(  1 + {\rm erf} \left( \frac{ v - \vmin}{\sqrt{2} \sigma_{v,\alpha}} \right)  \right) \right]^{-1}.
\end{align}

Studies such as \cite{2014A&A...562A..91P,2019arXiv190102016D} used bootstrapping methods to take into account the error distributions. They resampled the stars within their error bars to quantify the errors on the final values of the escape velocity. In this paper, we account for individual errors on all stars in the likelihood function. Although our method is more computationally intensive, we forward model all errors to obtain posterior distributions. A similar treatment of the errors was used by \cite{2020arXiv200616283K} (with the difference that ours includes an outlier model and a second component). 

Beyond the bound component, we also expect a small fraction of the stars to be either ejected or on unbounded orbits (e.g. \cite{2018ApJ...866..121H}). To account for such stars, we use an outlier model similar to that of \cite{2017MNRAS.468.2359W}, where
\be \label{eq:outlier}
p_{\alpha}^{\rm{out}}(\vobs) = \mathcal{A}  \exp \left( - \frac{\vobs^2}{2 [\sigma_{\rm{out}}^2 + \sigma_{v,\alpha}^2 ] } \right).
\ee
Unlike \cite{2017MNRAS.468.2359W}, which fixed the value of $\sigma_{\rm{out}} = 1000$ km/s, we marginalize over the dispersion $\sigma_{\rm{out}}$ of the outlier model as well as its fraction $f$. We also add in quadrature the measurement error of a particular star, although we expect it to be subdominant to $\sigma_{\rm{out}}$. We then normalize \Eq{eq:outlier} over the data region $[\vmin, \infty]$, and obtain 
\be
\mathcal{A}^{-1} = \sqrt{\frac{\pi}{2}} \sqrt{\sigma_{\rm{out}}^2 + \sigma_{v,\alpha}^2} ~\textrm{erfc} \left( \tfrac{\vmin \sqrt{\sigma_{\rm{out}}^2 + \sigma_{v,\alpha}^2} }{\sqrt{2}}  \right).
\ee

The likelihood per star $\alpha$ for a single bound component is therefore 
\be \label{eq:single_likelihood}
\mathcal{L}_\alpha^1 = (1-f) \tilde p_\alpha(\vobs^\alpha | \vesc, k) + f p_{\rm{out}}(\vobs^\alpha| \sigma_{\rm{out}}), 
\ee
while for two bound components it is
\begin{align}
\mathcal{L}_\alpha^2 =& (1-f) \left[ \,  f_S \tilde p_\alpha(\vobs^\alpha | \vesc, k_S)  \label{eq:double_likelihood} \right. \\
& \left. + (1-f_S) \tilde p_\alpha(\vobs^\alpha | \vesc, k) \, \right] + f p^{\rm{out}}_{\alpha}(\vobs^\alpha | \sigma_{\rm{out}}), \nonumber
\end{align}
where the slopes of the components are $k$ and $k_S$, and the fraction of the second component is labeled as $f_S$. This can be generalized to $n$ components.
The total log likelihood given by 
\be
	 \log \mathcal{L}^{i} = \sum_\alpha \log \mathcal{L}_\alpha^{i},
\ee
with $i = \{1,2\}$ the number of bound components assumed in the analysis. In what follows, we will refer to the two analyses as the ``single component" and ``two component" fits, by which we are discussing the bound components.  
In this work, we adopt the same terminology as in \cite{2019ApJ...874....3N}, where we call the relaxed component the Halo, and the kinematic substructure the Sausage. We emphasize that we do not know {\emph{a priori}} which component corresponds to which value of $k$.

We use the Markov Chain Monte Carlo \textit{emcee} \citep{2013PASP..125..306F} to find the best fit parameters, using 200 walkers, 500 steps for the burnin stage, and 2000 steps for each run. We next describe the parameters and priors.

\subsection{Priors}
\label{sec:priors}

\begin{table}[t]
\begin{tabular}{|l| c | c |}
\hline 
Parameter & Prior Range & Prior Type \\
 \hline 
 \hline 
$\vesc$ & $[\vmin, 1000]$ km/s & $1/v$ \\
$k$ & [0.1, 15] & Linear \\
$f$ & [$10^{-6}, 1$] & Log \\
$\sigma_{\rm{out}}$ & [3, 3000] km/s & Log \\
 \hline
 $k_S$ & [0.1,  $k$] & Linear \\
 $f_S$ & [0, 1] & Linear \\ 
 \hline
\end{tabular}
\caption{\label{tab:prior_1} List of the priors used in the analysis. }
\end{table}

The parameters of the single-component fit are  the escape velocity $\vesc$, the slope $k$, the fraction of the outlier distribution $f$, and the dispersion of the outliers $\sigmaOut$. For the two-component likelihood function, \Eq{eq:double_likelihood}, we add the slope of the second component, $k_S$, and its relative fraction with respect to the bound components, $f_S$. Without loss of generality, we assume $k_{S} < k$, but we remain agnostic as to the physical interpretation of each component.

We list these parameters in \Tab{tab:prior_1} along with their priors. The theory prior on the escape velocity is taken to be uniform in $1/\vesc$. Note that other authors such as \cite{2019arXiv190102016D} have taken a slightly different prior that is uniform in $\log \vesc$; given the narrow posteriors we will obtain, this choice will not impact results significantly. The fraction and dispersion of the outlier  distribution are taken to have log priors, given the large ranges that they might span. The fraction of the second component $f_{S}$ is taken to be linear  in $[0,1]$. The slopes $k$ and $k_S$ are taken to have linear priors. It important to emphasize that the default prior on the slopes is taken to be very wide,\footnote{We have also verified that our results are unchanged with the prior $k, k_S \in [0.1, 20]$} where $k, k_S \in [0.1, 15]$, as our goal to avoid shaping the distribution with restrictive priors. 

\subsection{Errors on speed measurements}
\label{sec:errors}

In this work, we will perform analyses on mock data generated with:
\begin{itemize}
\item \textbf{No Errors}: This will only be used on mock data to disentangle the effects of the errors from the rest of the pipeline. As expected, we cannot use this type of analysis for \Gaia data, as all data includes measurement errors. 
\item \textbf{Percentage Errors}: We assume the measurement errors on stars are a percentage of their true velocity. More explicitly, for each star with a true speed $\speed$, we sample its observed speed from a Gaussian distribution with a mean $\speed$ and a dispersion $x \times \speed$. Here $x$ is the percentage error, and we will consider as a representative value of $x = 5\%$. This case is closest to the \Gaia data, discussed in more detail in \cite{data_escape}. 
\end{itemize}
In addition, in the Appendix, we provide the results for mock data generated with the same absolute error in km/s for all stars, taking a representative value of 20 km/s since it is the most similar to the errors on stars in the \Gaia data. We find the conclusions are not sensitive to this particular choice. We also show results for the analysis with errors of 10\%, further discussed in \Sec{sec:other_cases}. 

\subsection{Akaike information criterion}
\label{sec:aic}

We will run the pipeline of \Sec{sec:standard} on a single component fit, as is standard in the literature, as well as the two component fit. In order to compare the two fits, we compute the Akaike Information Criterion (AIC) of each distribution, where the AIC is defined as \citep{aic}
\be
{\rm{AIC}} = 2 s - 2 \log(\hat{\mathcal{L}}), 
\ee
where $s$ is the number of parameters of the fit, and $\log(\hat{\mathcal{L}})$ is the maximum log likelihood of the fit. We compare the AIC of the single and two component fits to the data, where the one with the lower value of AIC is the better fit. Alternative functions, for example the Bayes Information Criterion (BIC) can also be used (see e.g. \cite{BIC}), but the AIC provided the most robust results when applied to simulations. The difference is in the way that it penalizes the added number of parameters, where AIC penalizes the models as  $2 s$ while BIC penalizes them as $s \log (n)$, with $n$ the total number of data points in the set \citep{vrieze2012model}. 

In this analysis, we will be evaluating 
\be \label{eq:delta_aic}
\Delta \rm{AIC}  = \rm{AIC}_2 - \rm{AIC}_1,
\ee
where AIC$_i$ is the AIC of the single ($i=1$) or double ($i=2$) component fit.

\section{Results with Simulated data}
\label{sec:simulations}

\begin{figure*}[t] 
   \centering
	\includegraphics[width=0.45\textwidth]{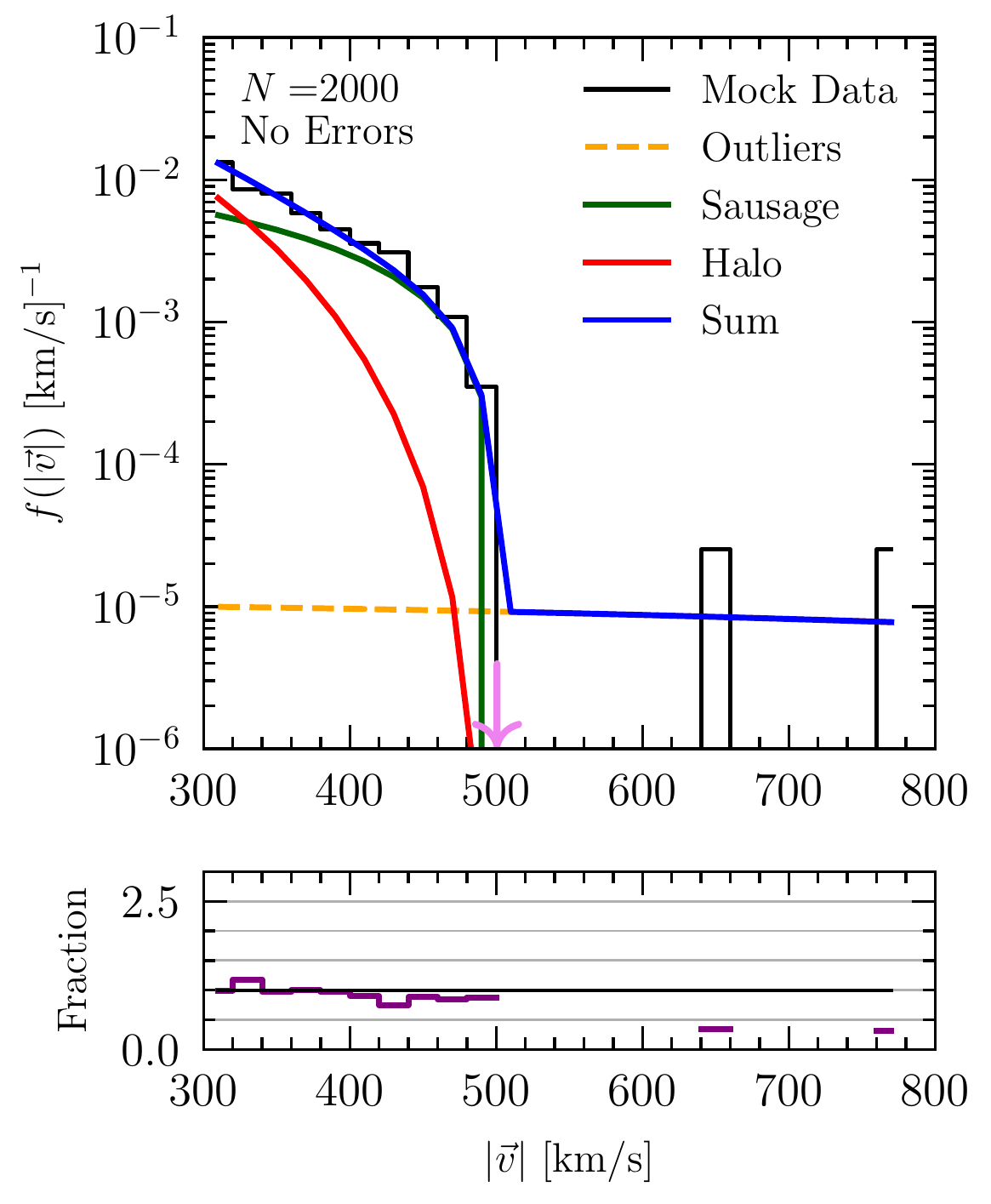} 
	\includegraphics[width=0.45\textwidth]{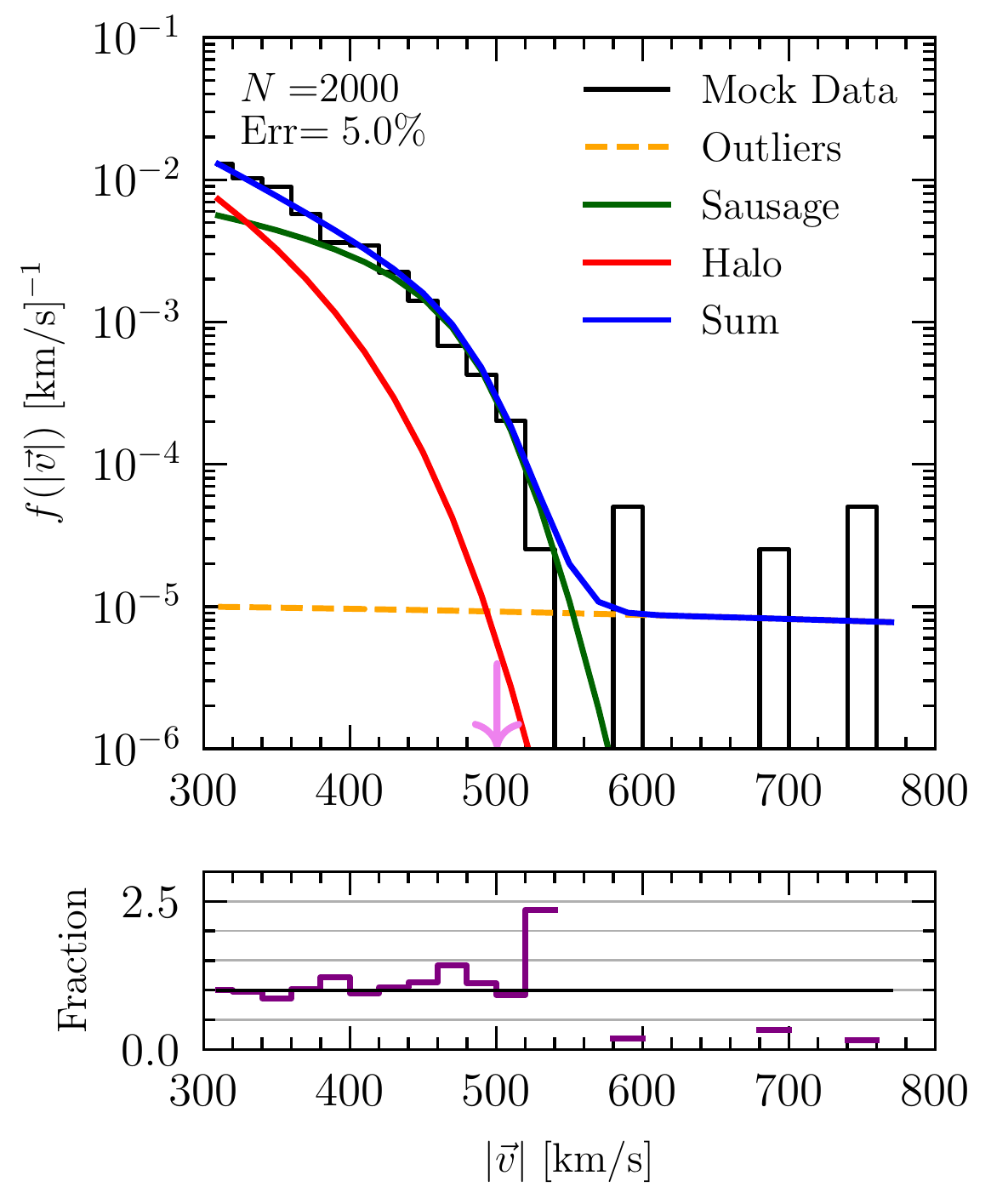} 
   \caption{(\textbf{Top}) Mock data drawn from a mixture of two bound components with $\vesc = 500$ km/s, and an outlier component with a fraction of 0.01 and a dispersion $\sigma_{\rm out}=1000$ km/s. The halo component has $k = 3.5$ and the substructure fraction is $f_S = 0.6$, with $k_S = 1.0$. The colored curves are the true distributions. (\textbf{Left}) We assume no measurement errors. (\textbf{Right}) We assume a measurement error of 5\% on the speeds. The true escape velocity is shown as a pink arrow. (\textbf{Bottom}) Ratio of the generated distribution to the true distribution. The generated stars follow the true distributions (accounting for errors), with fluctuations at high speeds due to the small outlier fraction. The missing bins are due to the lack of data in these bins.
   \label{fig:schematic_plot}
   }
\end{figure*}

We now present fit results analyzing mock data that includes a Halo component, a component due to a Sausage-like merger, and an outlier distribution. We thus explore how well the true $\vesc$ can be recovered in a fit, depending on different choices for $\vmin$, on the number of bound components in the fit, and on the priors for the slopes.

Throughout this section, we work with a fiducial sample of 2000 mock stars. The number of stars was chosen to be comparable to that found in the \Gaia data sample, and we assume $\vesc = 500$ km/s, with the slopes $k = 3.5$ and $k_S = 1.0$. The fraction attributed to the Sausage is $f_S = 0.6$ for $\vmin = 300$ km/s. The outliers are sampled from a Gaussian distribution with zero mean, a dispersion $\sigma_{\rm{out}} = 1000$ km/s, and an associated fraction $f = 0.01$.  From the fiducial sample, we generate three different data sets, by resampling each star from a Gaussian distribution with a mean given by the true speed and a dispersion given by its error. We will consider the two types of errors discussed in \Sec{sec:errors}: the no error sample, and the percentage error sample with errors of 5\%.  (An absolute error sample with errors set at 20 km/s for all stars is shown in Appendix~\ref{app:20kms_errors}.)

We show the fiducial sample of stars in \Fig{fig:schematic_plot}, where in the left panel there are no measurement errors and in the right panel we include percentage errors of 5\%.  The pink arrow shows the true escape velocity at 500 km/s, but in the right panel, the tail of the distribution extends out beyond 500 km/s due to the errors. It is thus imperative that the likelihood takes into account the presence of such errors. The ``true" distributions (solid curves) in the right panel are different from those on the left panel because we plot \Eq{eq:tail_error} instead of \Eq{eq:fvtail} in order to account for the presence of errors. 

From  \Fig{fig:schematic_plot}, we can immediately see that if the minimum velocity of the data sample $\vmin$ is too low, we might see more than a single distribution in the fit. The total distribution (blue) is dominated by the distribution with the lower $k$ for high enough speeds ($\speed \gtrsim $ 400 km/s). Below these values, the presence of the second distribution starts to dominate and will affect the fit. Using a single distribution would not produce the correct fit and slope, as we will explicitly show. Nevertheless, in order to have sufficient statistics and to obtain a reasonable fit to the distribution, previous works\footnote{Earlier studies performed this analysis with just the line-of-sight velocity measurements, so in order to have a fair comparison, we only compare to \cite{2018A&A...616L...9M,2019arXiv190102016D} as they have used three-dimensional velocity measurements.} have used rather low values of $\vmin = 250, 300$ km/s and obtained a local escape velocity $\vesc \sim 520-580$ km/s \citep{2018A&A...616L...9M,2019arXiv190102016D}.   It is not known {\emph{a priori}} where the cut should be, such that a single power law distribution is valid. Therefore, in this paper, we will use different values of $\vmin$ on mock data to show how this can provide an additional handle on the robustness of the result.

\begin{figure*}[t] 
   \centering
	\includegraphics[width=0.95\textwidth]{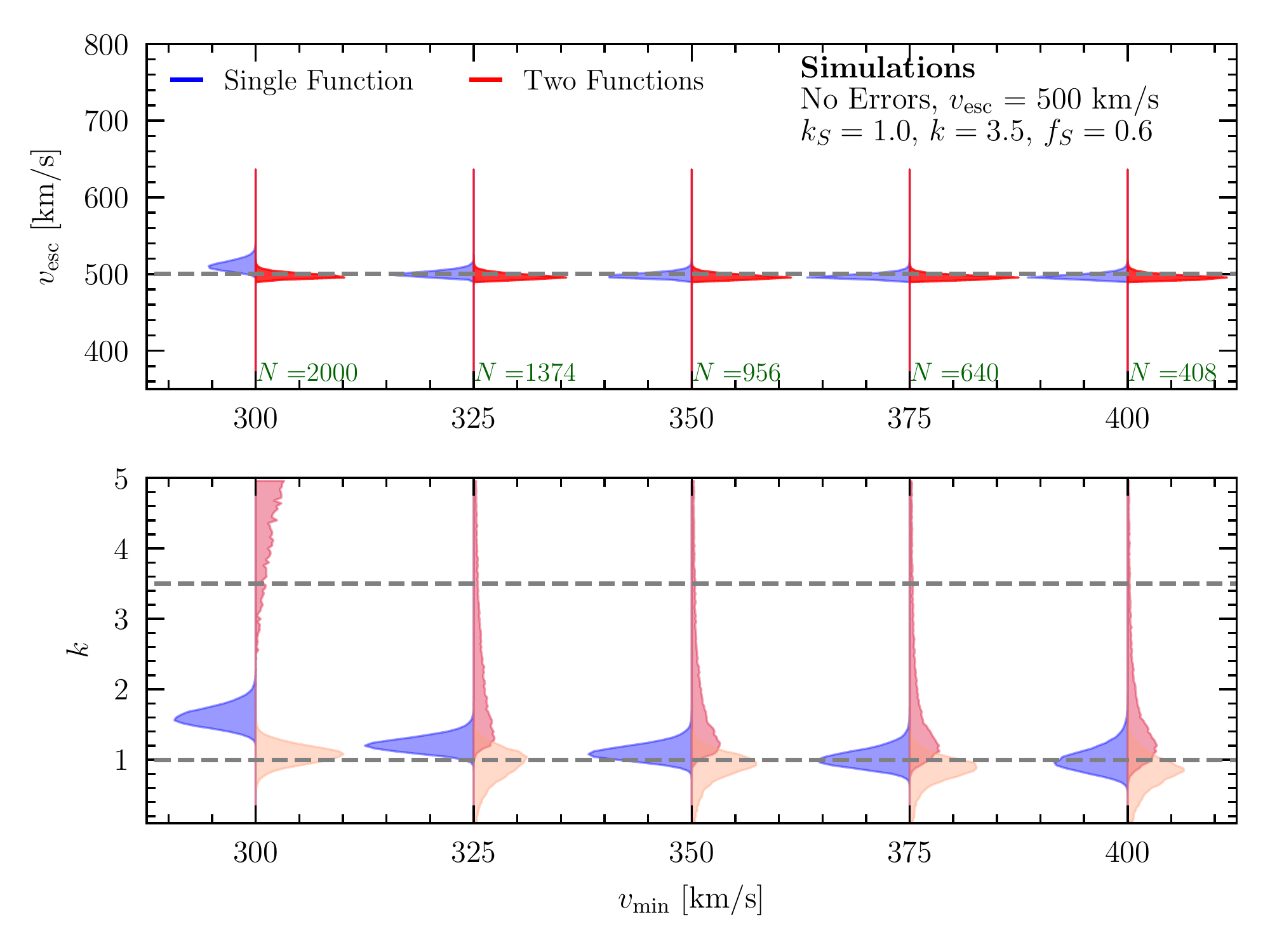} 
   \caption{Analyses of generated data with no errors, assuming the fiducial 2-component data set used throughout this section. ({\bf Top}) For each $\vmin \in [300,325,350,375,400]$ km/s, we show on the vertical axis the posterior distribution of $\vesc$  for a single-component fit (blue) and the two-component fit (red). The true value is indicated by the dashed grey line.  ({\bf Bottom}) Posterior distribution of $k$ for a single-component fit (blue) and of $k_{S}$ (lighter red) and $k$ (darker red) for the two-component fit.
   }
   \label{fig:mock_sims_no_err}
\end{figure*}

We now proceed by implementing the analysis outlined in \Sec{sec:analysis} for different sets of minimum velocities, with $\vmin \in [300, 325, 350, 375, 400]$ km/s. We do not generate a separate data set for each run, but rather use the exact same data set throughout, which leads to the number of stars per sample decreasing as the minimum velocity changes. For example, for the sample with no errors, the number of stars in the sample is 2000 stars for $\vmin = 300$ km/s, while it drops to 408 stars for $\vmin = 400$ km/s. The \Gaia data behaves similarly, thus we can account for the effect of decreasing statistics with increased cutoff velocity in this manner.

\subsection{No Error Analysis}
\label{sec:no_errors}

We first assume perfect measurements while analyzing generated data. Doing so helps validate the pipeline and isolate the effects of the errors. In \Fig{fig:mock_sims_no_err}, we show the posterior distributions of $\vesc$ and of the slope(s) for different values of $\vmin$. We show both the single component fit (\Eq{eq:single_likelihood}), as has been previously implemented, as well as the two component analysis (\Eq{eq:double_likelihood}).  The dashed gray lines in \Fig{fig:mock_sims_no_err} are the true values.

\begin{figure*}[t] 
   \centering
      \includegraphics[trim={0 0 0 0},clip,width=0.45\textwidth]{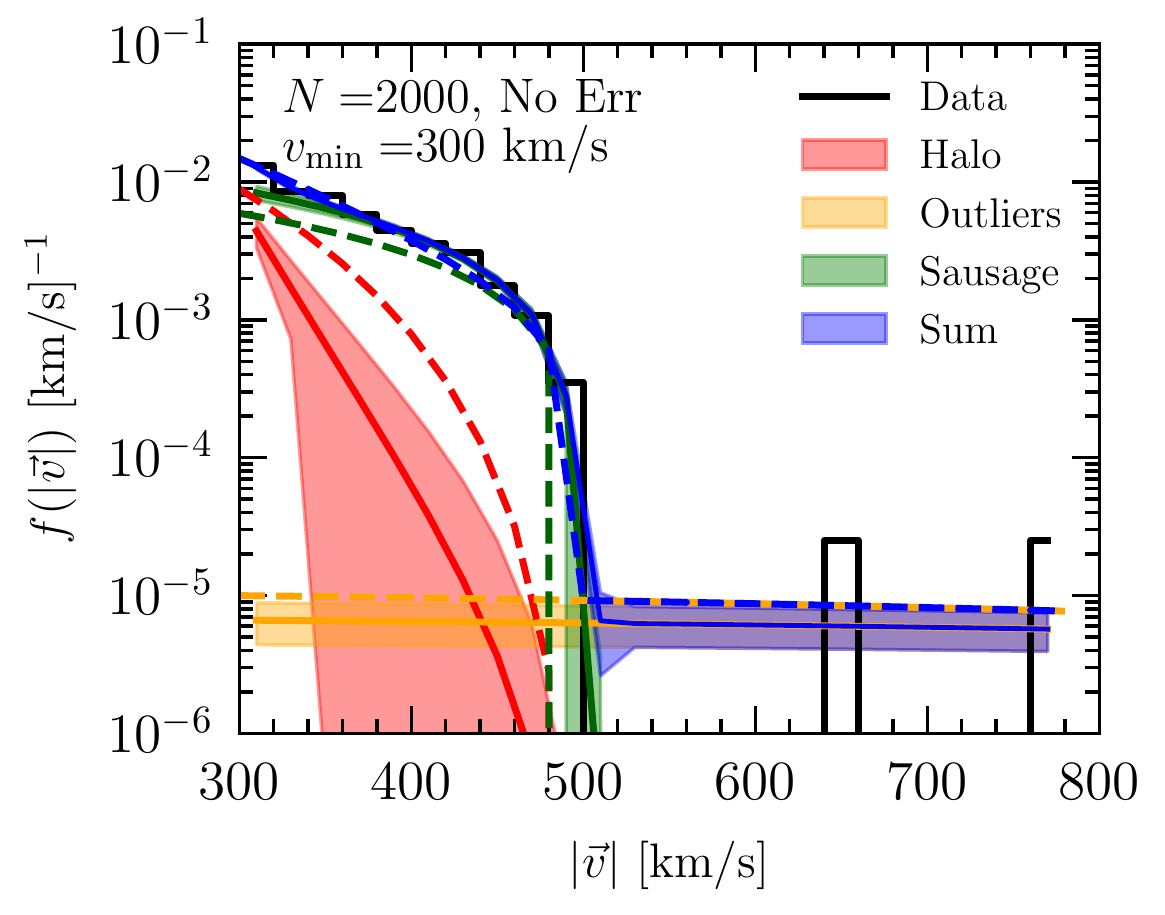} 
	\includegraphics[trim={0 0 0 0},clip,width=0.45\textwidth]{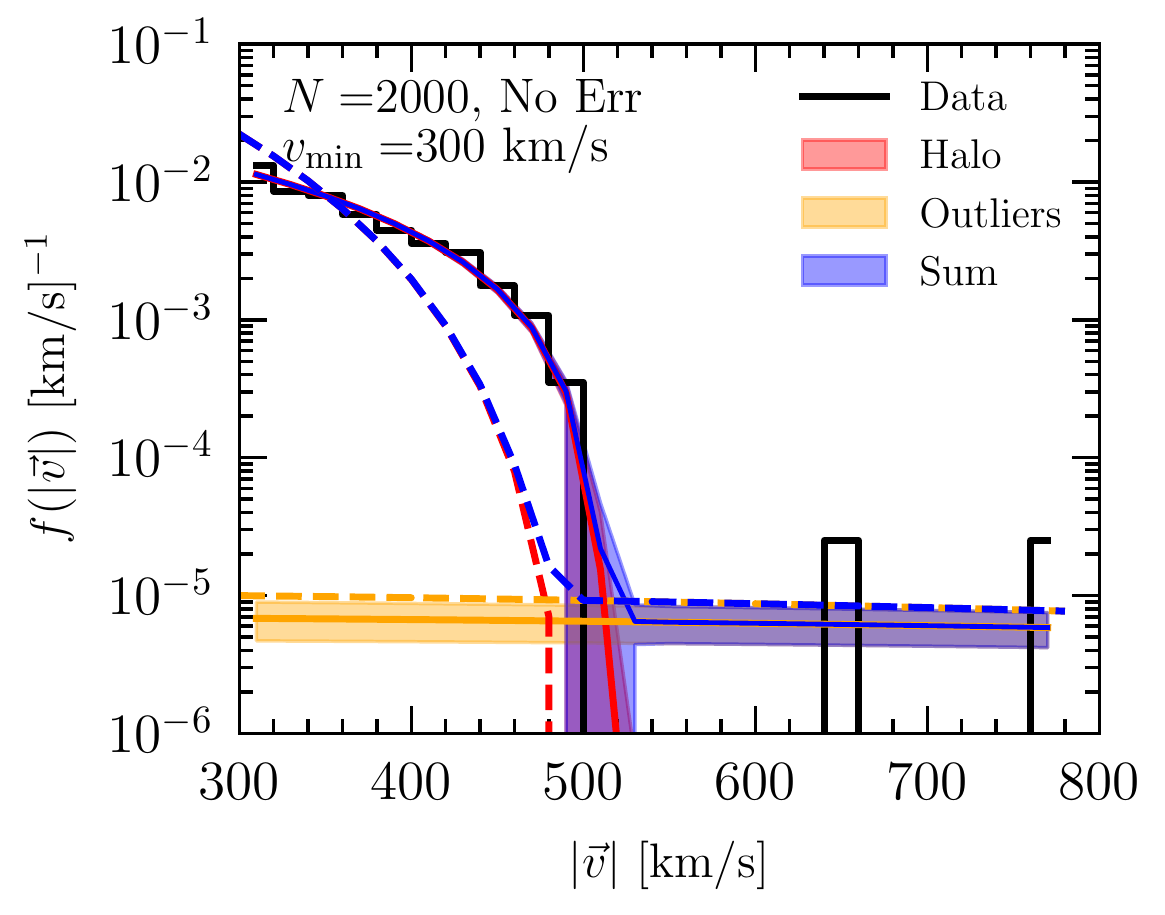} 
   \caption{Best fit speed distributions overlaid on a histogram of mock data, assuming perfect measurements. 
   ({\bf Left}) Best fit distributions assuming two function fit; the true distributions are shown in dashed lines, and the best fit in solid lines. The shaded regions are $68\%$ containment regions obtained from the posteriors for each component. The distributions shown are the Halo (red), the Sausage (green), the outliers (yellow), and the summed distribution (blue). ({\bf Right}) Best fit distributions for a single function fit. The functions shown are the Halo (red) and outliers (yellow). The summed distribution is shown in blue. The ``true" distribution shown in this case is that of the Halo alone, normalized to the full distribution. }
   \label{fig:mock_overfit_no_errors}
\end{figure*}

We begin by considering the sample with $\vmin > 300$ km/s, and 2000 stars. We find that the two component fit, in the absence of errors, accurately obtains the correct escape velocity and substructure slope. More explicitly, the best fit values for the escape velocity and the slope of the Sausage are $\vesc = 497.46^{+3.98}_{-2.84}$ km/s and $k_S = 1.07^{+0.13}_{-0.13}$, both of which are within a standard deviation of the true values of $\vesc = 500$ km/s and $k_S = 1$ (see corner plot in \Fig{fig:corner_plot_no_error}).  Meanwhile, the posterior distribution of $k$ for the Halo extends to high values, $k = 8.98^{+3.48}_{-2.93}$, which is only within two standard deviations of the true value. The high slope might lead us to incorrectly conclude that a single function fit is sufficient. However, doing so in this case anchors the single component to a slope larger than that of the Sausage, resulting in an overestimate of the escape velocity with $\vesc = 510.71^{+7.20}_{-5.81}$ km/s and with $k = 1.61^{+0.15}_{-0.13}$ (full corner plot can be found in \Fig{fig:corner_plot_no_error_1component}).

To understand these fit results in terms of the speed distributions, we show these resulting range of distributions in \Fig{fig:mock_overfit_no_errors} for the two component analysis (left panel) and the single component analysis (right panel). 
The shaded regions are  the 68\% containment regions for each component, while the dashed lines are the true model distributions.
In the left panel, we see that the range of Halo distributions does not match the true Halo component, and has quite a large uncertainty. Disentangling that particular slope is quite difficult given the narrow range of values in which it dominates ($\speed \in [300, 350]$ km/s, as can be seen in the true distributions). From the corner plot in \Fig{fig:corner_plot_no_error}, we see that there is a degeneracy between the substructure fraction and the Halo slope, where a larger Halo slope can be compensated by increasing $f_{S}$. The fit, however, is finding the correct model of the tail of the speed distribution, and subsequently the correct escape velocity. The outlier distribution is also correctly recovered.

In the right panel of \Fig{fig:mock_overfit_no_errors}, we overlay the fit results for the single component analysis. With a single function, the fit has to account for both the lower speed stars near $\speed \sim 300$ km/s with a larger $k$, as well as the slope and cutoff near $\vesc$. Doing so with a single function leads to overestimating the slope (since it is averaging the slopes of the two distributions) as well as the escape velocity. Because of the limited statistics of stars near $\vesc$, the fit is largely influenced by the distribution at lower speeds. This is why it is imperative to understand the physics of the objects being modeled when extracting a physical quantity such as $\vesc$. 

Returning to the results with larger $\vmin$ in \Fig{fig:mock_sims_no_err}, we find that as $\vmin$ increases, the single function and two function fits quickly converge on the correct value. This is the expected behavior since the speed distribution for larger $\speed$ is then dominated by a single function.  The example here illustrates that if there are kinematic substructures that not captured by the model, this would lead to results that drift with $\vmin$. If the un-modeled components peak at particular speeds, this could also lead to results that are not even monotonic with $\vmin$. \textit{Redoing the fit for increasing values of $\vmin$ is therefore a consistency test of the fit.}  

\subsection{Percentage Errors}
\label{sec:percent_errors}

\begin{figure*}[t] 
   \centering
	\includegraphics[width=0.95\textwidth]{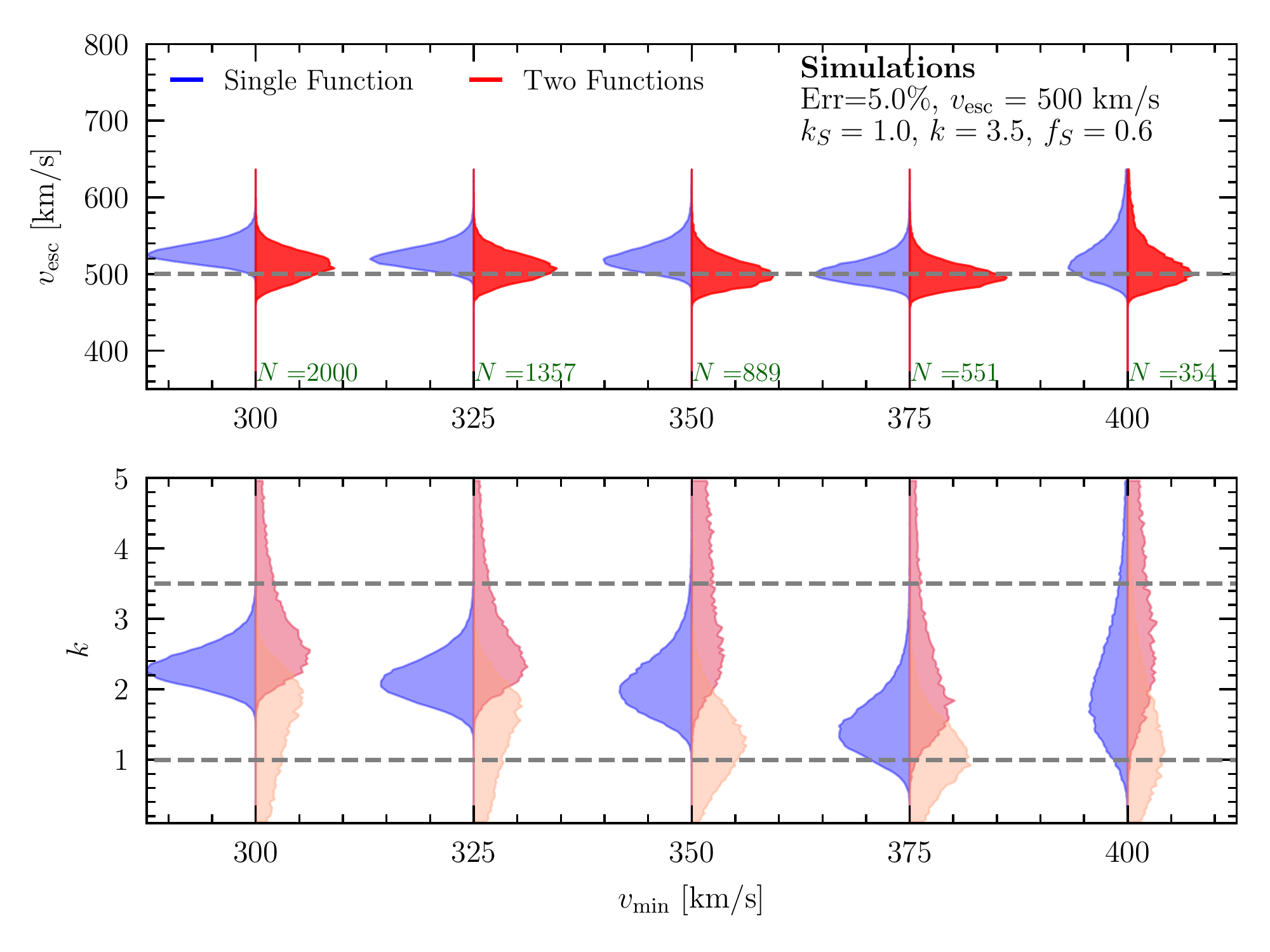} 
   \caption{Similar to \Fig{fig:mock_sims_no_err}, with the stellar speeds now sampled from a Gaussian distribution with a dispersion of $5\%$ of the true speed.  }
   \label{fig:mock_sims_percent}
\end{figure*}

We next repeat the analysis using percentage errors of 5\% on the speed of each star (as described in \Sec{sec:errors}). This case is most similar to what we expect to find in the \Gaia data. We show the results of the fit as a function of $\vmin$ in \Fig{fig:mock_sims_percent}. The results are qualitatively similar to the case of no errors, but with larger spread in the posterior distributions. In the single component fit with $\vmin = 300$ km/s, both $\vesc$ and the substructure slope $k_S$ are biased towards larger values, with best fit values $\vesc = 527.60^{+14.40}_{-12.00}$ km/s and $k = 2.32^{+0.29}_{-0.24}$. This is again due to the fact that the slopes of the two distributions tend to get averaged with a single fit, while the two component fit accounts for the presence of the substructure.  

We find that even including realistic observational errors, the two-component fit can obtain robust fits to $\vesc$ that are consistent with the true value for all $\vmin$. For example, in the case of $\vmin = 300$ km/s, $\vesc = 511.43^{+17.94}_{-18.42}$ km/s. The slopes' posteriors are $k = 2.97^{+2.05}_{-0.64}$, and $k_S = 1.60^{+0.57}_{-0.86}$,  consistent with the true values within a single standard deviation. In fact, the Halo slope is better constrained in the presence of errors than without. This difference might just reflect shot noise in the generated distributions, which get smeared out more when observational errors are included.

\begin{figure}[t] 
   \centering
      \includegraphics[trim={0 0 0 0},clip,width=0.45\textwidth]{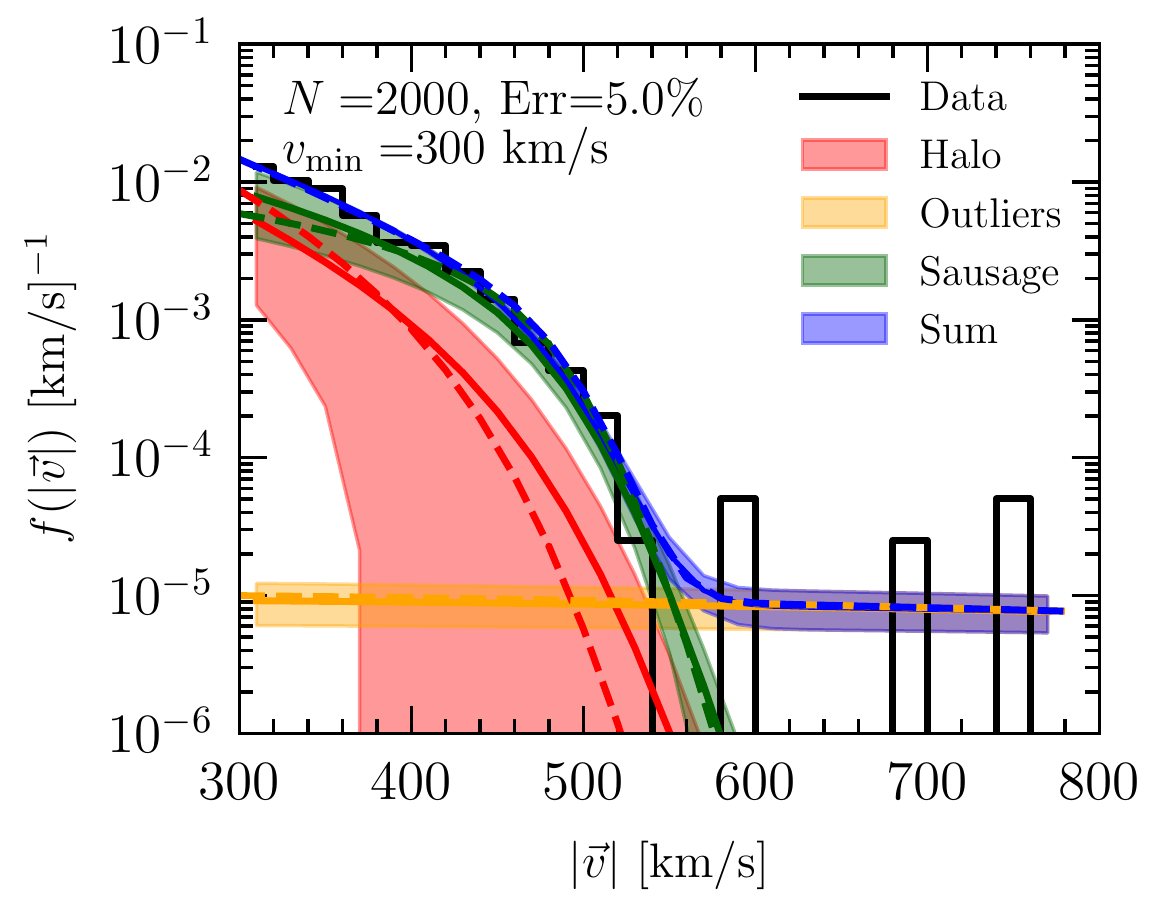} 
   \caption{Similar to the left panel \Fig{fig:mock_overfit_no_errors}, we show the best fit speed distributions overlaid on a histogram of the mock data.  Here we take stellar speeds sampled from Gaussian distributions with 5\% measurement error. Dashed lines are true distributions, accounting for errors.}
   \label{fig:mock_overfit_5pct_error}
\end{figure}

In \Fig{fig:mock_overfit_5pct_error}, we plot the best fit distributions from the two-component analysis against the data points. Note that in plotting the best fit (solid lines) and true distributions (dashed lines), we have taken into account the presence of errors. To do so, we average over the errors of the stars in the sample, and use \Eq{eq:tail_error} with these values. We find that the Halo component (red) is poorly constrained, similar to \Fig{fig:mock_overfit_no_errors}. This is again because the Halo component dominates for only a small range of velocities. The full distribution, however, is a good fit to the data. 

\begin{figure}[t] 
   \centering
      \includegraphics[width=0.45\textwidth]{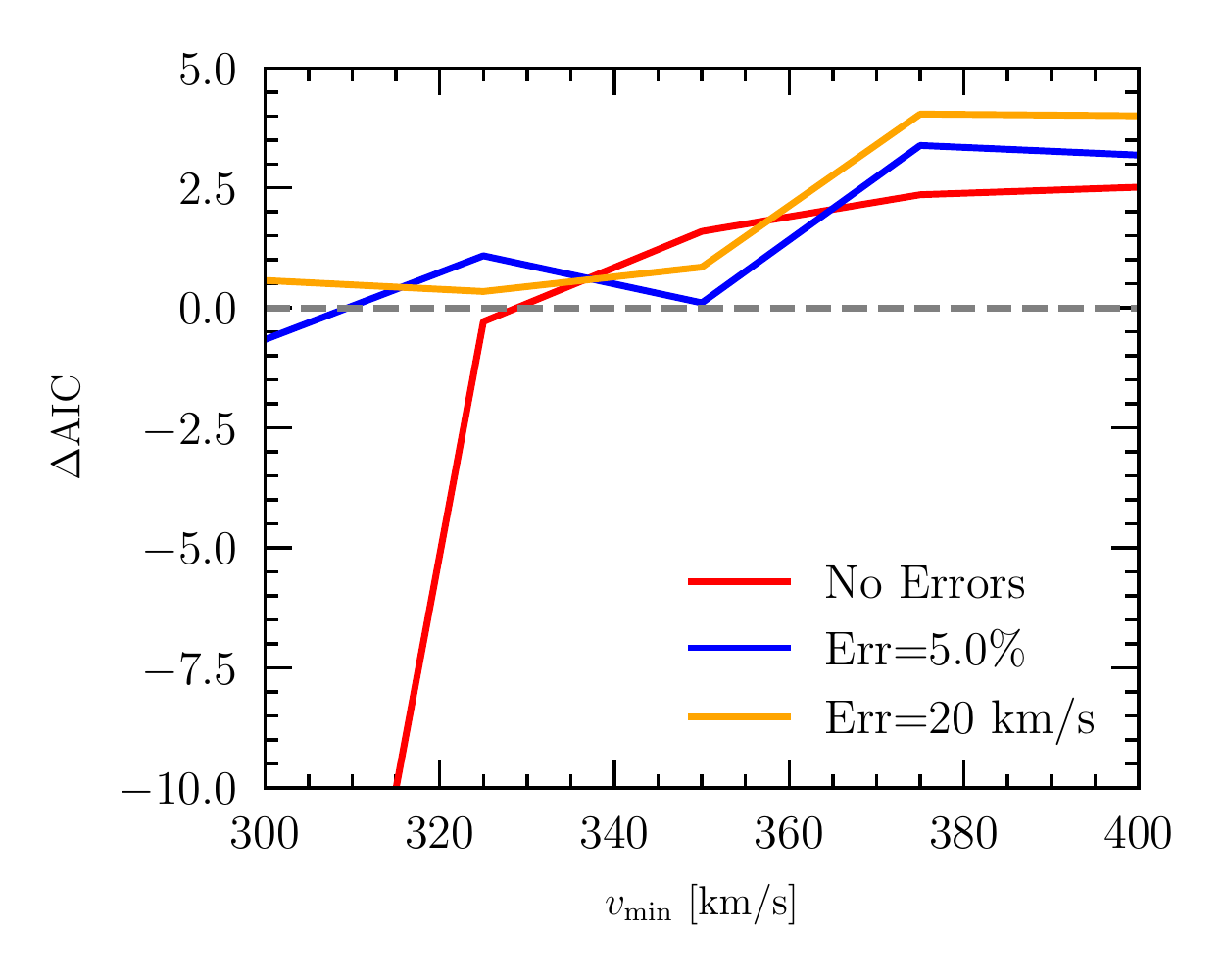} 
   \caption{$\Delta$AIC for the two-component fit compared with the one-component fit. Negative values indicate the two-component fit is favored. For larger $\vmin$, only a single function is needed and the $\Delta$AIC reflects the penalty for introducing additional parameters.}
   \label{fig:mock_aic}
\end{figure}

For a more quantitative analysis of the goodness of fit, we evaluate the $\Delta$AIC, introduced in \Sec{sec:aic}, for the different values of $\vmin$. We compare the two function fit to the single function fit using \Eq{eq:delta_aic}. Negative values of the $\Delta$AIC correspond to a better fit for the two component model over the single component model. We show the resulting $\Delta$AIC in \Fig{fig:mock_aic} for the analyses with and without errors (\Sec{sec:no_errors} and \Sec{sec:percent_errors}).  In the absence of observational errors, we find that the two component fit is overwhelmingly preferred for the lower values of $\vmin$. For higher  $\vmin \ge 350$ km/s, the one-component fit can capture the tail well, and we see $\Delta$AIC tends towards positive values since there is a penalty for extra model parameters. From \Eq{eq:delta_aic}, we would expect $\Delta$AIC = 4 if the maximum likelihoods were exactly equal. 

In the presence of 5\% errors on the speeds, the $\Delta$AIC does not favor either the single or two-component fit at lower $\vmin$. At larger $\vmin$ and with 5\% errors, the $\Delta$AIC again tends towards positive values favoring the single-component fit, as physically expected since a single distribution dominates. A similar behavior is shown for runs with an absolute error of 20 km/s, which are discussed further in the Appendix in \Sec{sec:absolute_errors}.

Although the goodness of fit does not show a preference for the two-component analysis, it is important to note that this analysis does show an escape velocity that is robust to varying $\vmin$. This can be seen in \Fig{fig:mock_sims_no_err} and \Fig{fig:mock_sims_percent}. For example, for the two component fit for the benchmark data set with 5\% errors, the best fit at $\vmin = 300$ km/s is $\vesc = 511.43^{+17.94}_{-18.42}$ km/s, and at $\vmin = 400$ km/s is $\vesc = 509.18^{+35.42}_{-20.29}$ km/s.  This is not true for the single component fit, where the results drift with $\vmin$. In this case, for $\vmin = 300$ km/s, the best fit is $\vesc = 527.60^{+14.40}_{-12.00}$ km/s, while for $\vmin = 400$ km/s it is $\vesc = 519.18^{+36.31}_{-22.06}$ km/s. \textit{This means that one has to check the goodness of fit for a different number of components, but also test the robustness of the results as a function of $\vmin$, a strategy we adopt when studying the \Gaia DR2 results in \cite{data_escape}.}

\subsection{Effect of Limited Priors}
\label{sec:limited_prior}

To compare with the standard analysis in the literature, we next show the result of a single component fit with limited priors on $k$. Such priors have been used before to deal with limited data samples and the degeneracy in the fits for $\vesc$ and $k$, but leads to different results for $\vesc$ depending on the prior chosen.

\begin{figure}[t] 
   \centering
	\includegraphics[width=0.48\textwidth]{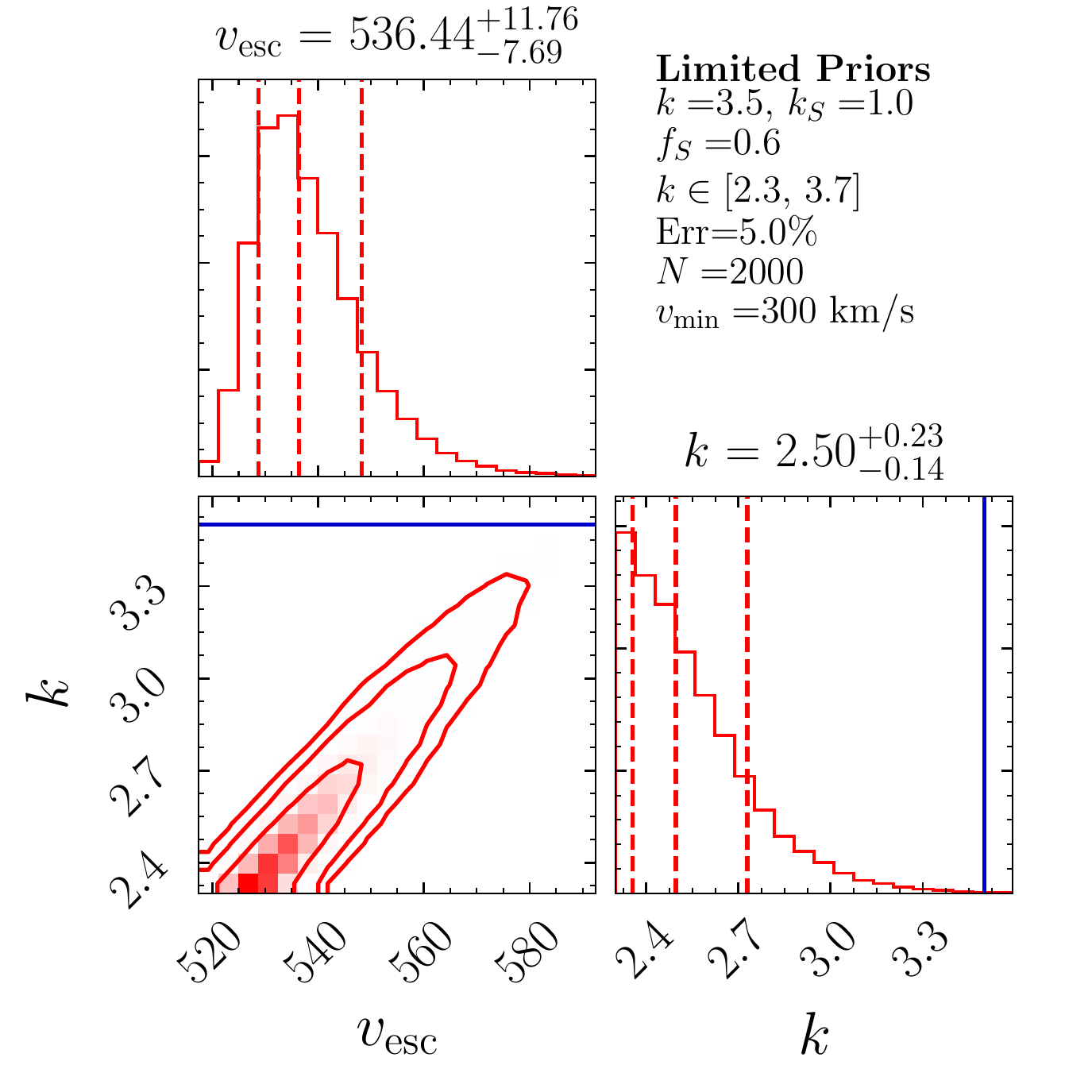} 
   \caption{Using the same data set from \Fig{fig:mock_sims_percent} with $\vmin = 300$ km/s, we perform a single-component analysis with a prior on the slope $k \in$ [2.3, 3.7]. This mimics the setup of \cite{2014A&A...562A..91P}. In the presence of substructure, $\vesc$ is biased towards larger values. }
   \label{fig:limited_prior}
\end{figure}

Here we use the same data set as described above with $\vmin = 300$ km/s and $5\%$ errors, and now impose the hard prior of \cite{2014A&A...562A..91P}, where $k \in [2.3, 3.7]$. Note that the halo component's slope $k = 3.5$ is within the range of the priors, but the second slope $k_S = 1.0$ is not. The results are shown in \Fig{fig:limited_prior}. The effect of the limited priors is to bias $\vesc$ towards even larger values than in the analysis with loose priors, shown in  \Fig{fig:mock_sims_percent}.  This again reflects the fact that the single function fit will tend to find an averaged slope of the two components, with a tight correlation in $\vesc$ and $k$. A strong prior on $k$ towards larger values will then further bias $\vesc$. Since we do not {\emph{a priori}} know the slopes and sizes of the individual components, placing strong priors on $k$ can lead to incorrect inferences about $\vesc$.  As the statistics and quality of the data improve, this effect would likely become obvious if the posteriors are seen to pile up along the edge of the $k$ prior. 

\subsection{Estimating the Substructure Fraction and Slope}

\begin{figure}[t] 
   \centering
	\includegraphics[width=0.45\textwidth]{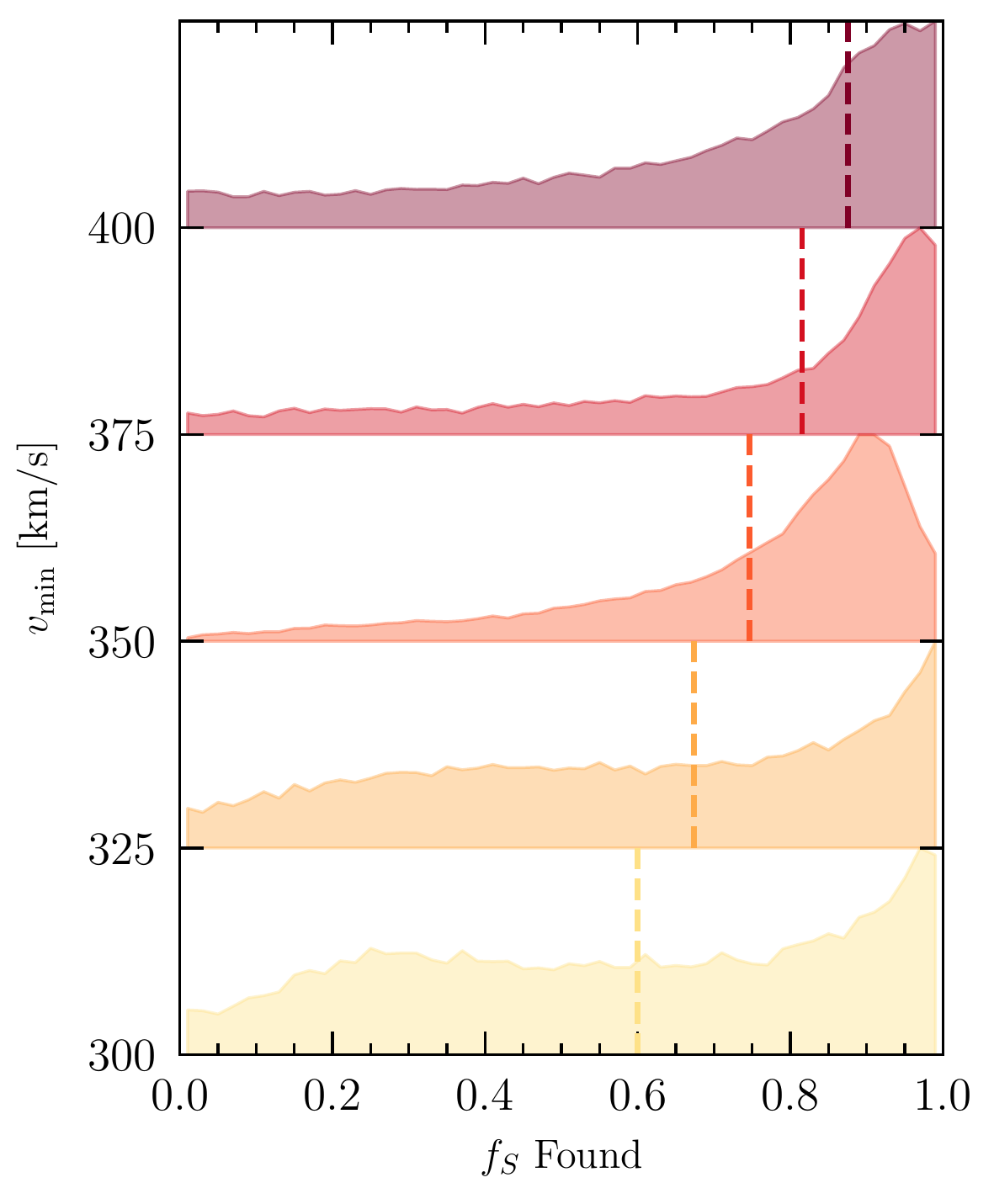} 
   \caption{True versus recovered substructure fraction posteriors from the analysis of \Sec{sec:percent_errors}, where we assumed 5\% errors on the measured speeds of the stars. The true fractions are shown as vertical dashed lines, and are ordered from lightest to darkest corresponding to the values of $\vmin = [300, 325, 350, 375, 400]$ km/s. }
   \label{fig:frac_sims_percent}
\end{figure}

It is interesting to see how well we can reconstruct properties of the substructure component, as it would offer independent information on the Milky Way merger history from other studies. Based on \Fig{fig:mock_sims_percent}, it is possible to constrain the slope $k_{S}$, where in that analysis we obtained $k_{S} = 1.60 ^{+0.57}_{-0.86}$ for $\vmin = 300$ km/s, for a true value of $k_S = 1$. As pointed out in \cite{2019arXiv190102016D}, this slope is correlated with the assembly history and can be compared with the predictions of cosmological simulations for different merger mass and time. Although it is difficult to constrain the larger of the slopes, the smaller one is well constrained and robust throughout our analyses, as shown in \Fig{fig:mock_sims_percent} (and \Fig{fig:mock_sims_absolute} for the example with absolute errors). In general, this smallest slope could likely be attributed to the Sausage based on the reasoning of \cite{2019arXiv190102016D}.

In addition, one of the parameters that we marginalize over is the fraction of the non-outlier distribution associated to the Sausage, $f_{S}$. Note that the parameter $f_{S}$ in a given analysis is not exactly the same as the value of $f_{S} = 0.6$ used in generating the mock data set. This is because the fraction changes as a function of $\vmin$, which can be seen for example in \Fig{fig:schematic_plot}. We thus compute the true values of $f_S(\vmin)$ by integrating the true distributions in the interval $[\vmin, \infty]$ km/s as shown in \Eq{eq:fraction}. The distributions $\mathbf{f_S}$ and $\mathbf{f_H}$ are now those with the average errors convolved in them: {\emph{i.e.,}} the distributions in the right panel of \Fig{fig:schematic_plot}. 

In \Fig{fig:frac_sims_percent}, we  show the recovered posteriors of the fraction $f_S$ along with the true values for $f_S(\vmin)$, indicated by dashed vertical lines.  The distributions are shaded from lightest to darkest as $\vmin$ increases from 300 km/s to 400 km/s. We find that the posteriors of the fractions are not properly converged, and it is hard to extract the correct fraction of this distribution. This is because the Halo component is difficult to constrain when it dominates for only a narrow range of speeds. This highly depends on the differences between the slopes, and as we will see in \cite{data_escape}, the fractions are better constrained in the \Gaia DR2 analysis.  

\subsection{Additional Substructure}

Including the second component allows us to model and reconstruct a smoothly falling substructure component, but there might be even more kinematic substructures or speed distributions which are not well-described by a power law. Such features might also appear as a result of data selection cuts or kinematically incomplete samples. To test how this could affect our fits, we consider mock data sets injected with additional Gaussian distributions that are peaked at speeds above 300 km/s. This kind of substructure could be a small fraction of the overall stellar distribution, but still strongly influence fit results if it is peaked at large $\speed$. 

We consider two cases: (1) a Gaussian with mean of $370$ km/s, dispersion of 20 km/s, and fractional contribution $f_{i} = 0.2$, and (2) a Gaussian with mean of $500$ km/s, dispersion of 50 km/s, and fractional contribution $f_{i} = 0.01$. The first case introduces an additional feature at 370 km/s that cannot be entirely modeled with two power law distributions, while the latter case introduces a feature near $v_{\rm esc}$ that can easily lead to confusion between the tails of the Sausage/Halo and the outlier population. We created mock data sets assuming these Gaussians in addition to the Sausage, Halo, and outliers with fiducial parameters used throughout the section; the latter distributions were then weighted by $1 - f_{i}$. We take the same number of total stars as before, $N=2000$ for $\vmin = 300$ km/s, and we again assumed observational errors of 5\%. 

\begin{figure*}[t] 
   \centering
      \includegraphics[trim={0 0 0 0},clip,width=0.45\textwidth]{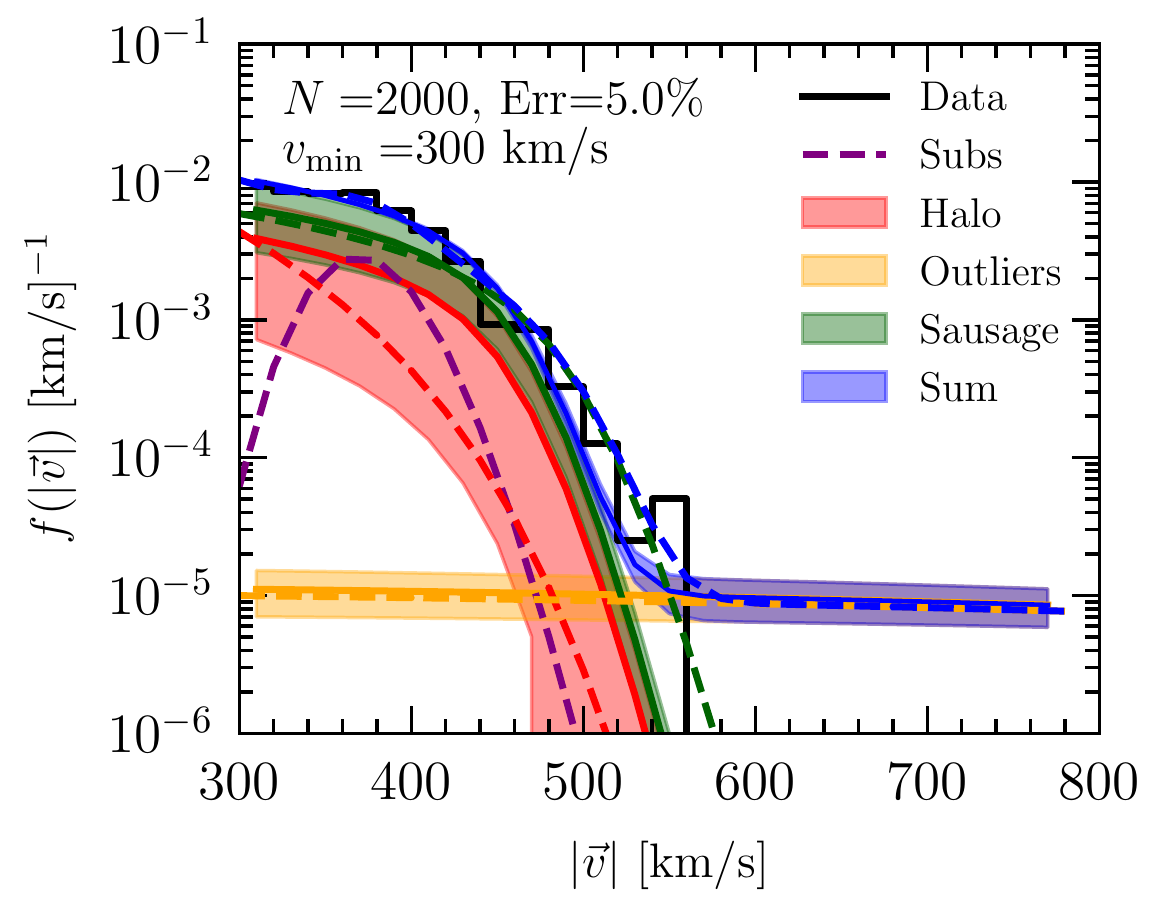} 
      \includegraphics[trim={0 0 0 0},clip,width=0.45\textwidth]{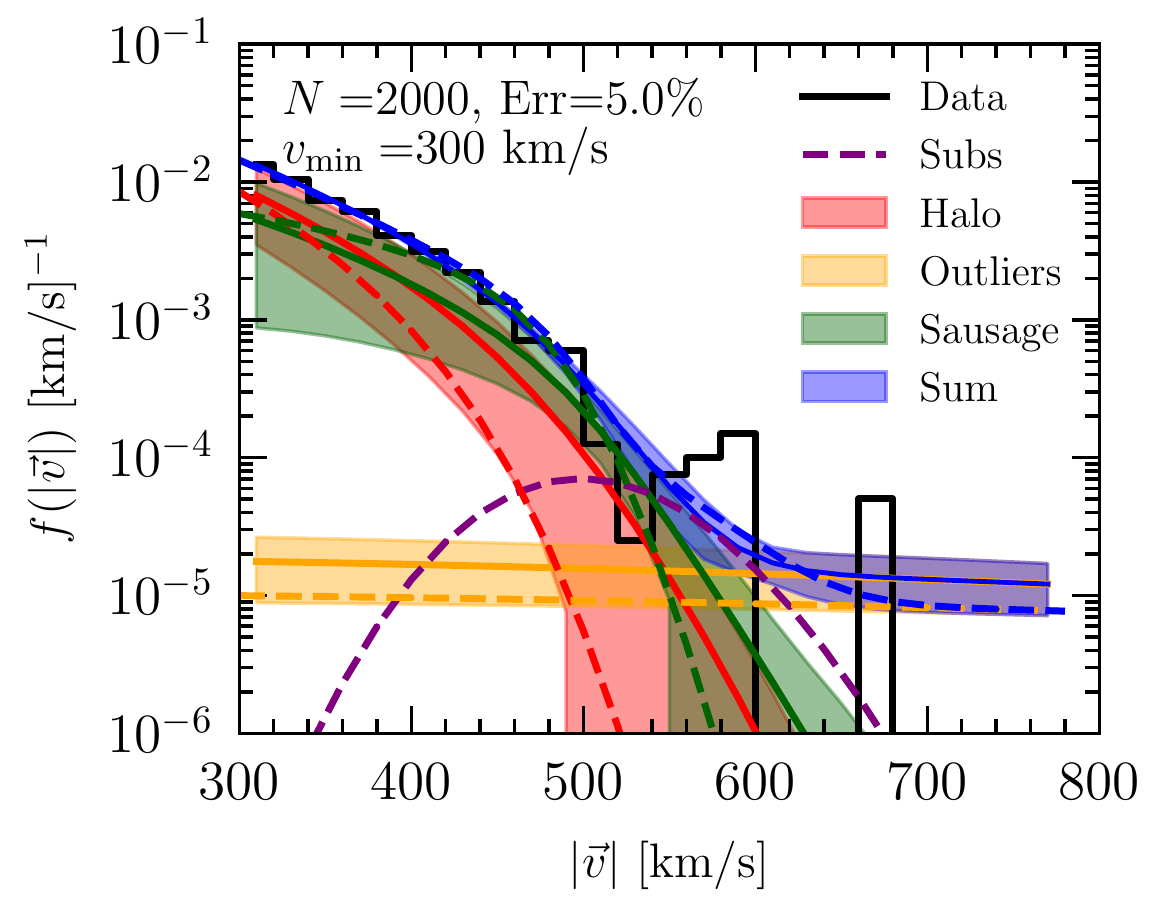} 
   \caption{Best fit distributions (shaded bands) over a histogram of the mock data, where the mock data set includes an additional injected Gaussian feature. ({\bf Left}) We inject a Gaussian centered at 370 km/s, with a dispersion of 20 km/s, and a fraction $f_i = 0.2$, which leads to an additional feature in the speed distribution instead of a smoothly falling function. In order to fit the feature, the slopes of the Halo and Sausage are driven to smaller values, as can be seen in Fig.~\ref{fig:mock_vesc_k_gaussian_370kms} for $\vmin = 300$ km/s.  ({\bf Right})  We inject a Gaussian centered at 500 km/s with a dispersion of 50 km/s and a fraction $f_i = 0.01$, which smears the tail of the velocity distribution and drives the fit towards larger $\vesc$. }
   \label{fig:mock_overfit_gaussian}
\end{figure*}

\begin{figure*}[t] 
   \centering
	\includegraphics[width=0.95\textwidth]{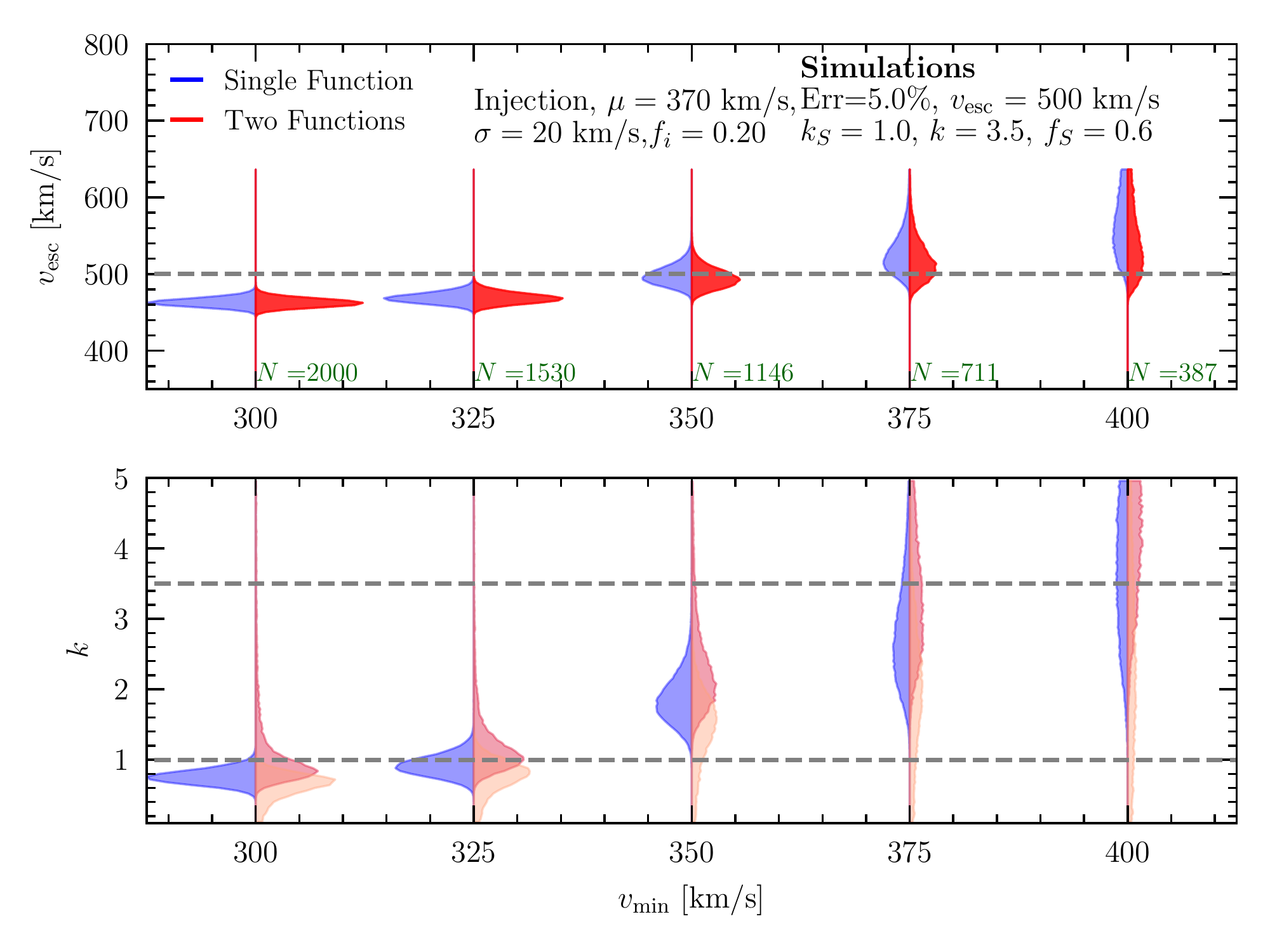} 
   \caption{Fit results as a function of $\vmin$ for a mock data set where we injected an additional Gaussian substructure component with mean 370 km/s, dispersion 20 km/s, and fractional component $f_{i}$ = 0.2. A drift in $\vesc$ for even the two-component cases suggests additional unmodeled structure at lower $v$.  }
   \label{fig:mock_vesc_k_gaussian_370kms}
\end{figure*}

Histograms of the mock data sets are shown in Fig.~\ref{fig:mock_overfit_gaussian}, along with the true distributions (dashed lines). The shaded bands are the best fit distributions from a two-component fit. In the first case (left panel), the Gaussian at 370 km/s introduces an additional bump which flattens out the speed distribution below 400 km/s. The best fit slopes are correspondingly driven to low values in order to fit this feature, with $k_{S} = 0.65^{+0.17}_{-0.15}$. Due to the degeneracy between $k$ and $\vesc$, this leads to $\vesc$ being significantly underestimated with $\vesc= 462.4^{+5.8}_{-5.4}$~km/s for $\vmin = 300$ km/s. Fig.~\ref{fig:mock_vesc_k_gaussian_370kms} shows the resulting posteriors for $\vesc$ and $k$ as a function of $\vmin$, for both single and two-component fits. In this case, the single and two-component fits give similar results at low $\vmin$, where the fit is driven strongly by the additional injected Gaussian. As we increase $\vmin$ to 375 km/s and above, we cut out most of the injected feature and the fits again converge to the correct $\vesc$. This example shows that both one- and two-component fits can exhibit $\vesc$ results that drift with $\vmin$ when there is peaked substructure between $\vmin$ and $\vesc$.

\begin{figure*}[t] 
   \centering
	\includegraphics[width=0.95\textwidth]{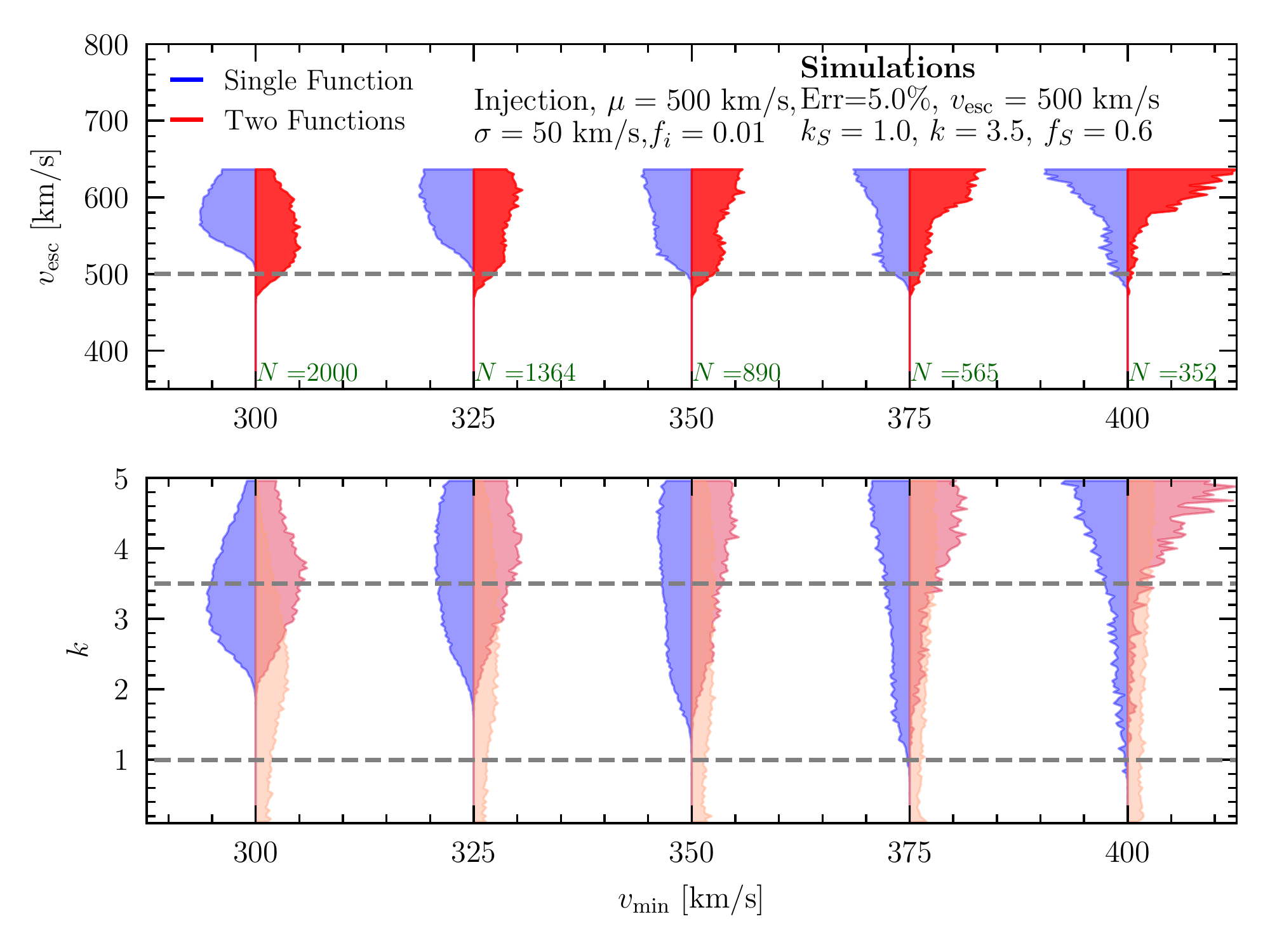} 
   \caption{Fit results as a function of $\vmin$ for a mock data set where we injected an additional Gaussian substructure component with mean 500 km/s, dispersion 50 km/s, and fractional component $f_{i}$ = 0.01. The sizeable degeneracy in $\vesc$ and $k$, as well as the non-convergent results at larger $\vmin$, are indicative of unmodeled structure near the tail or mismodeled outliers population. }
   \label{fig:mock_vesc_k_gaussian_500kms}
\end{figure*}

In the second case we studied, the additional Gaussian is near $\vesc$. This example illustrates what can happen in the fits if there is an additional population which is clustered near $\vesc$, or if the outlier population is not captured by our model. As shown in the right panel of Fig.~\ref{fig:mock_overfit_gaussian}, the best fit $\vesc$ is driven to larger values in order to capture this additional component, with $\vesc= 563^{+54}_{-46}$~km/s for $\vmin = 300$ km/s. The posteriors in $\vesc$ and $k$ as a function of $\vmin$ are shown in Fig.~\ref{fig:mock_vesc_k_gaussian_500kms}. We see that the presence of the additional unmodeled component near $\vesc$ leads to a much more severe degeneracy in $\vesc$ and $k$ for both single and two-component fits. Thus, seeing highly degenerate or non-convergent results even for high $\vmin$, where we expect a single component to dominate, may indicate the presence of additional high-speed substructure or mismodeled outlier population.

While the Gaussian injections considered here are artificial, they are useful to illustrate the effect in two somewhat extreme cases. These two cases show how additional unmodeled substructure may drive $\vesc$ systematically higher or lower, depending on where this substructure peaks. If the substructure peaks at lower $\speed$, this can be diagnosed if we obtain $\vesc$ results that depend on $\vmin$, and mitigated by selecting large enough $\vmin$.  If the fits exhibit a large amount of degeneracy and do not converge even with good statistics, it might be a sign that our outlier model is not sufficient. This case will be more challenging to get around. However, performing these tests can allow us to be more confident that the fit results are not being strongly driven by the presence of unmodeled components. 

\subsection{Further case studies}
\label{sec:other_cases}

We test the method introduced in this work in a number of other scenarios, which we summarize here. We leave the full details for \App{sec:app_checks}. 

{\emph{Similar slopes:}} If the values of the slopes $k$ and $k_S$ are similar, then it becomes more difficult to separate the two components. We test this case with an analysis on mock data with $k_{S} = 1$ and $k=2$, and show the results of the fits in  \Fig{fig:mock_sims_slopes}. With a low $\vmin$, it is still possible to extract the two slopes with the two-component analysis, but we find that it is difficult to extract the individual slopes for larger $\vmin$. However, the value of $\vesc$ remains robust to the change in the slopes. Furthermore, if this case exists in data, we can verify it by checking the behavior of the single and two component fits as a function of $\vmin$ and using the AIC. When the slopes are very similar, we expect to see a similar $\vesc$ result between the single and two-component fits at low $\vmin$. 

{\emph{Large errors:}} Another case we checked is if the error in the measurements are larger. In Appendix~\ref{app:large_errors}, we show the results of an analysis with 10\% errors on the speeds, keeping the fiducial model parameters the same as in the rest of this section. In this case, the tail of the speed distribution is shaped strongly by the error distribution and it is difficult to distinguish the two components or their slopes. Although the recovery of the slopes is difficult in this case, we find that the escape velocity obtained remains quite robust throughout the full analysis. Given that a realistic data sample has errors close to 5\%, this case can also be avoided with quality cuts on the stars. 

These examples further illustrate why it is valuable to perform single and multi component fits as a function of $\vmin$. By understanding the behavior in both cases, we can also infer whether the results may be biased by one of the limiting scenarios discussed here.

\section{Conclusions}
\label{sec:conclusions}

In order to extract sensible results for the local stellar escape velocity, recent studies have taken low values for $\vmin$, where it is not clear if the power-law distribution of \cite{1990ApJ} holds, and imposed artificial priors on the slope of the speed distribution $k$. These choices can shape the measurement of the escape velocity and the mass of the Milky Way. In this paper, we focused on building a robust strategy to obtaining the escape velocity, with results independent of the choice of the priors. Our pipeline accounts for individual errors on stellar speeds in a forward model, as well as the outlier distribution. Most importantly, we account for the presence of multiple kinematic substructure components in the speed distribution for the first time. 

A second kinematic component in the speed distribution is motivated by the presence in our local neighborhood of (at least) a second kinematic structure besides the stellar halo, called the \Gaia Sausage.
To account for this, studies of the tail of the speed distribution either need to increase the minimum velocity $\vmin$ above which we define a ``tail," or make sure that we have the correct number of components in the model. Thus including a second kinematic component is physically motivated. 

To model the presence of this substructure, we introduce a second bound component following \Eq{eq:fvtail}, with the same escape velocity but a new slope for the tail $k$. The approach can also be generalized to include more components. We then fit for the escape velocity, the slopes of the structures, and their fractional contributions. The fit is repeated with different numbers of components, and different definitions of the tail of the distribution (i.e., different values of $\vmin$). 

Using mock data, we found that our pipeline can reconstruct $\vesc$ in the presence of substructure, and furthermore is robust to changes in slope and the presence of observational errors. One lesson drawn from these results is that it is crucial to study the dependence of the fit results as we increase $\vmin$, the minimum speed for the data set.  We have shown how a single component fit could be biased for low $\vmin$, while at high $\vmin$ the result of the single and two component fits should converge. In parallel, we can also measure the AIC as a goodness of fit test, and check that it prefers a single function as $\vmin$ increases.  A strong drift in $\vesc$ and $k$ with $\vmin$ could be an indication that the model is missing an important component of the data.

To argue against using tight priors as has been the standard in the field, we also analyzed the mock data with the priors used in \cite{2014A&A...562A..91P}. When the slope of a single-component fit is limited by the priors, the inferred escape velocity will be strongly affected. The choice of priors in previous studies is based on simulations, and depends on the merger history of the simulations considered. Since we do not know {\emph{a priori}} what the true slopes should be and currently do not have empirical evidence on these values, we should keep the priors wide in order to obtain a robust measurement of the escape velocity.

In a companion paper \cite{data_escape}, we use the method outlined here to measure the escape velocity of the Milky Way. We apply the single and two function fits over the five values of $\vmin$, as was done on the mock data sets. There we show that a multicomponent fit does provide a better fit to \Gaia data, allowing us to extract a robust escape velocity of $\vesc = 484.6^{+17.8}_{-7.4}$~km/s and Milky Way mass of $M_{200} = 7.0^{+1.9}_{-1.2} \times 10^{11} M_{\odot}$.

\section*{Acknowledgements}

We are grateful to I. Moult for early discussions and collaboration on the project, and to M. Lisanti for helpful feedback.
We would also like to thank L. Anderson, A. Bonaca, G. Collin, A. Deason, P. Hopkins, A. Ji, and J. Johnson for helpful conversations. 

This work was performed in part at Aspen Center for Physics, which is supported by National Science Foundation grant PHY-1607611.
This research used resources of the National Energy Research Scientific Computing Center (NERSC), a U.S. Department of Energy Office of Science User Facility operated under Contract No. DE-AC02-05CH11231. 
LN is supported by the DOE under Award Number DESC0011632, the Sherman Fairchild fellowship, the University of California Presidential fellowship, and the fellowship of theoretical astrophysics at Carnegie Observatories.
TL is supported by an Alfred P. Sloan Research Fellowship and Department of Energy (DOE) grant DE-SC0019195.

\pagebreak

\def\bibsection{} 
\bibliographystyle{aasjournal}
\bibliography{v_escape}

\pagebreak
\clearpage
\appendix

\setcounter{equation}{0}
\setcounter{figure}{0}
\setcounter{table}{0}
\setcounter{section}{0}
\makeatletter
\renewcommand{\theequation}{S\arabic{equation}}
\renewcommand{\thefigure}{S\arabic{figure}}
\renewcommand{\theHfigure}{S\arabic{figure}}
\renewcommand{\thetable}{S\arabic{table}}

\FloatBarrier

\section{Results with absolute errors of 20 km/s \label{app:20kms_errors} }
\label{sec:absolute_errors}

In the main text, we generated mock data assuming observational errors of 5\%. Here we show results for an absolute error of 20 km/s. This leads to observational errors that are similar in magnitude, although smaller at high speeds, and the fit results are qualitatively similar as well.

\begin{figure*}[h] 
   \centering
	\includegraphics[width=0.95\textwidth]{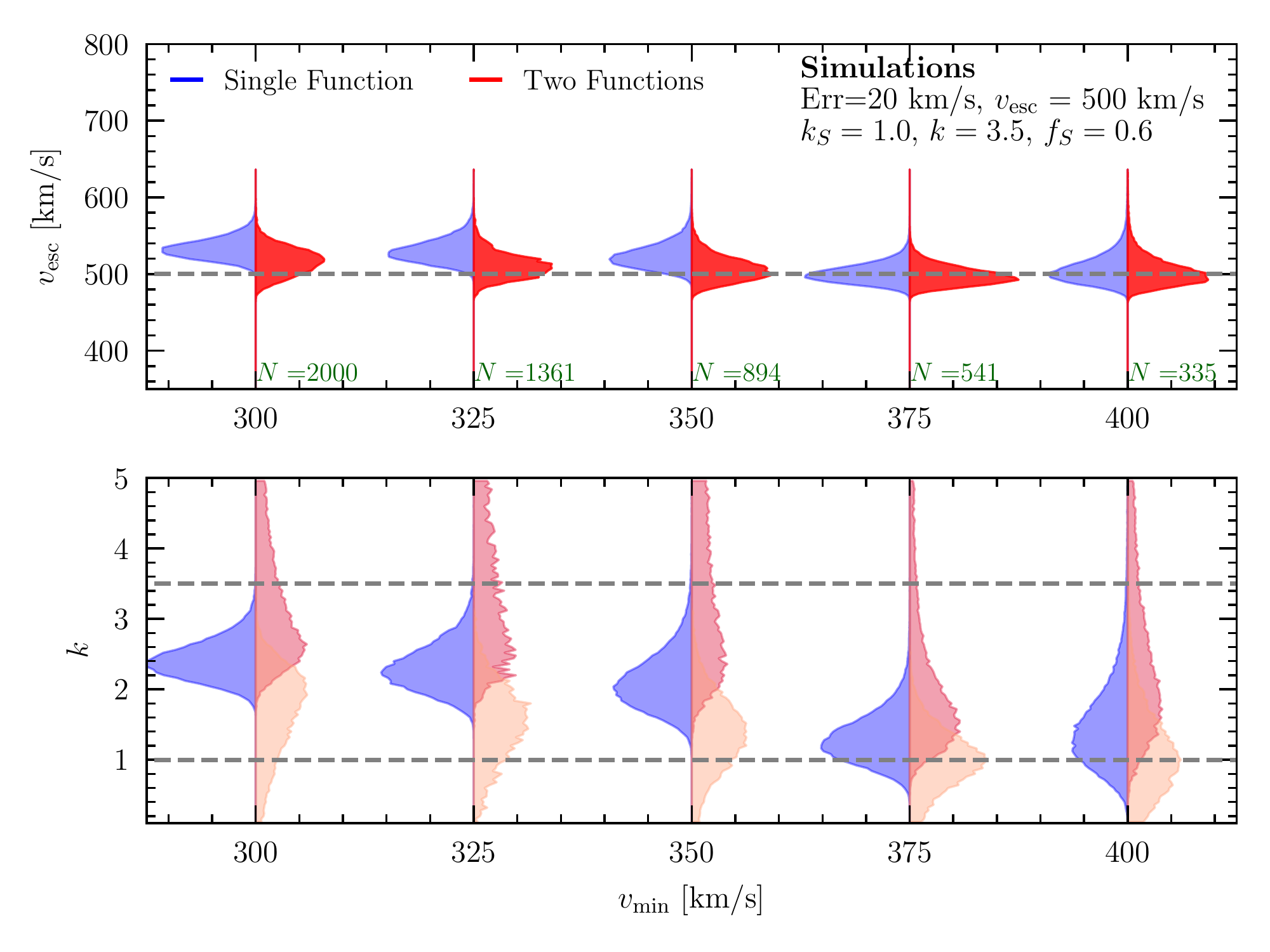} 
   \caption{Similar to \Fig{fig:mock_sims_no_err}, with the observed stellar speeds now sampled from a Gaussian distribution about the true speed with a dispersion of 20 km/s. The results are qualitatively similar to the case with 5\% errors.}
   \label{fig:mock_sims_absolute}
\end{figure*}

\begin{figure}[t] 
   \centering
      \includegraphics[trim={0 0 0 0},clip,width=0.45\textwidth]{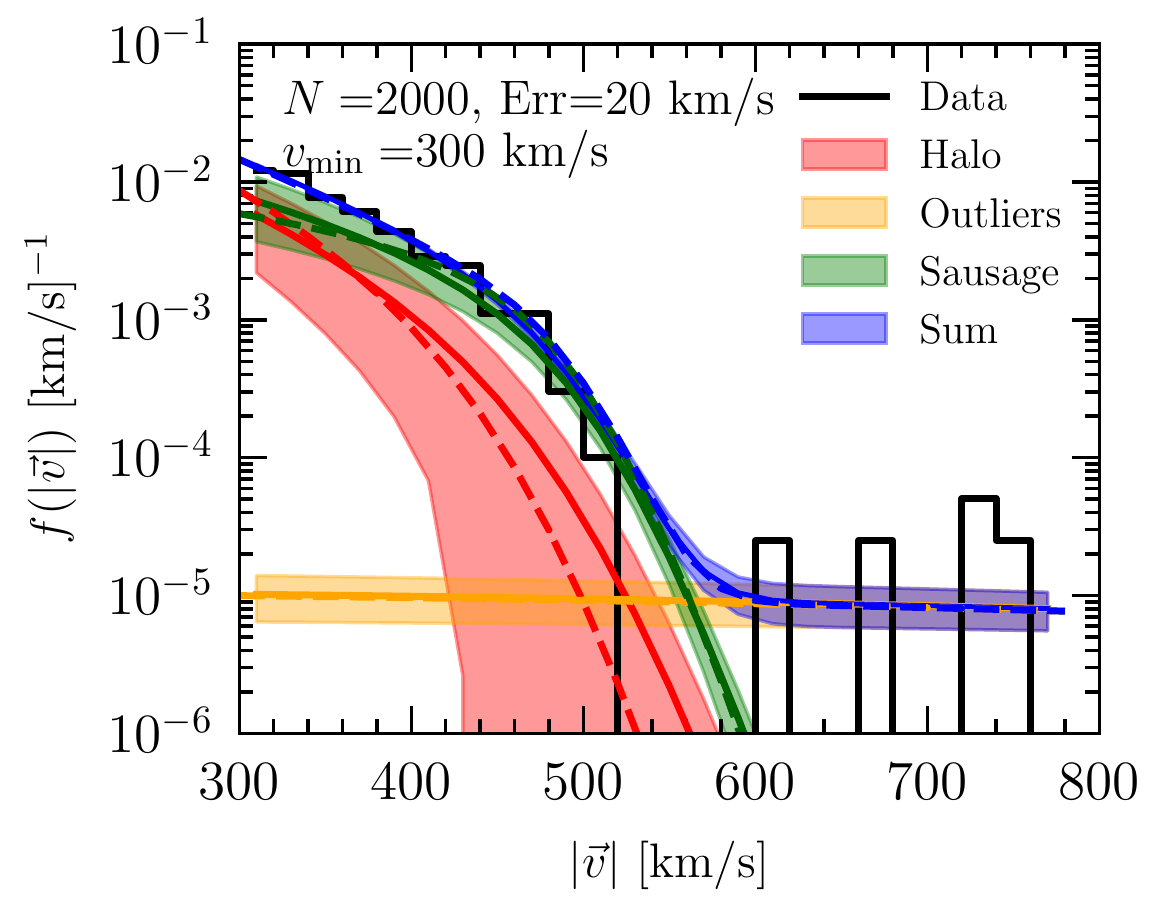} 
      	\includegraphics[trim={0 0 0 0},clip,width=0.45\textwidth]{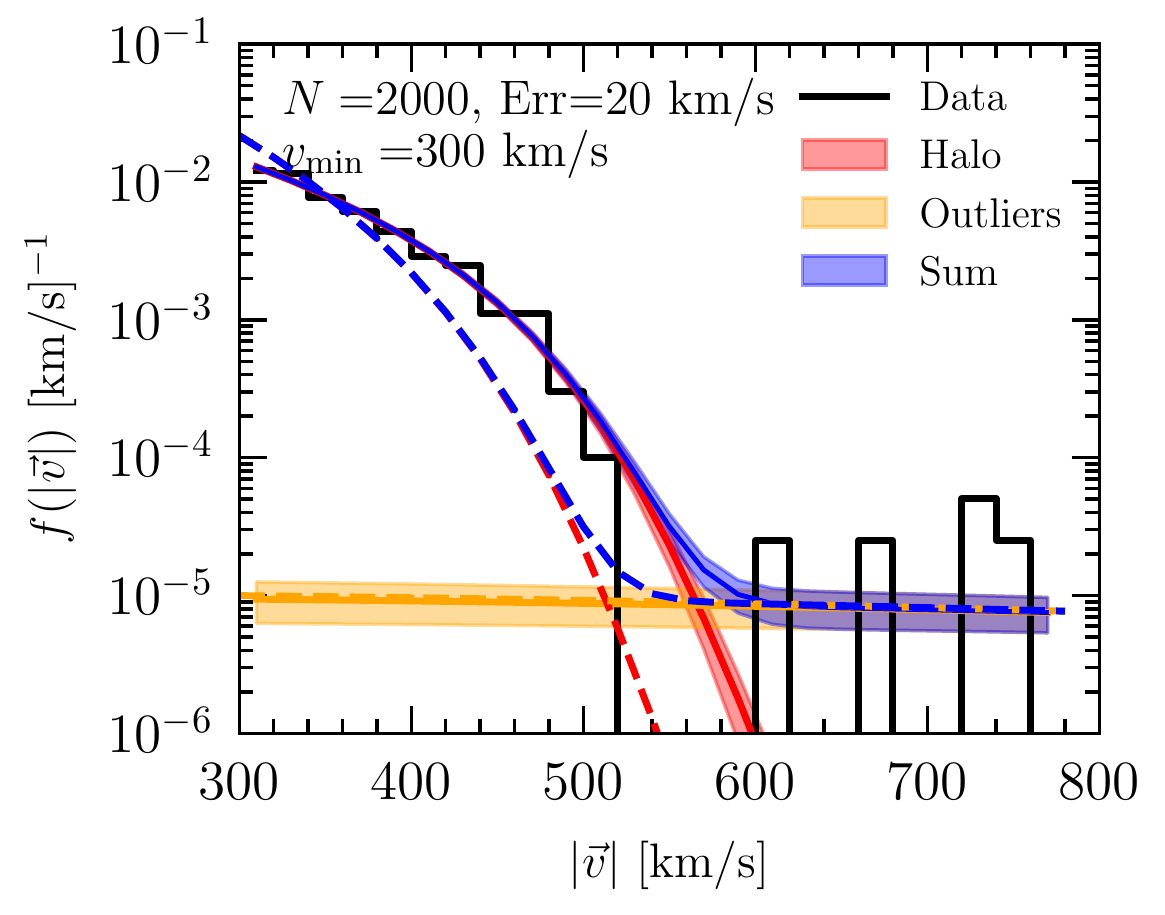} 
   \caption{Similar to the left panel of \Fig{fig:mock_overfit_no_errors}, with the errors on the stellar speed sampled from Gaussian distributions with 20 km/s dispersions. The single component fit (right panel) overestimates the escape velocity. }
   \label{fig:mock_overfit_20_error}
\end{figure}

\clearpage
\FloatBarrier

\section{Corner Plots}

In this section we show the corner plots of the fits of the main text. In addition, we present the corner plots for the case of 20 km/s errors.

\subsection{No errors}

\begin{figure}[h] 
   \centering
	\includegraphics[width=0.95\textwidth]{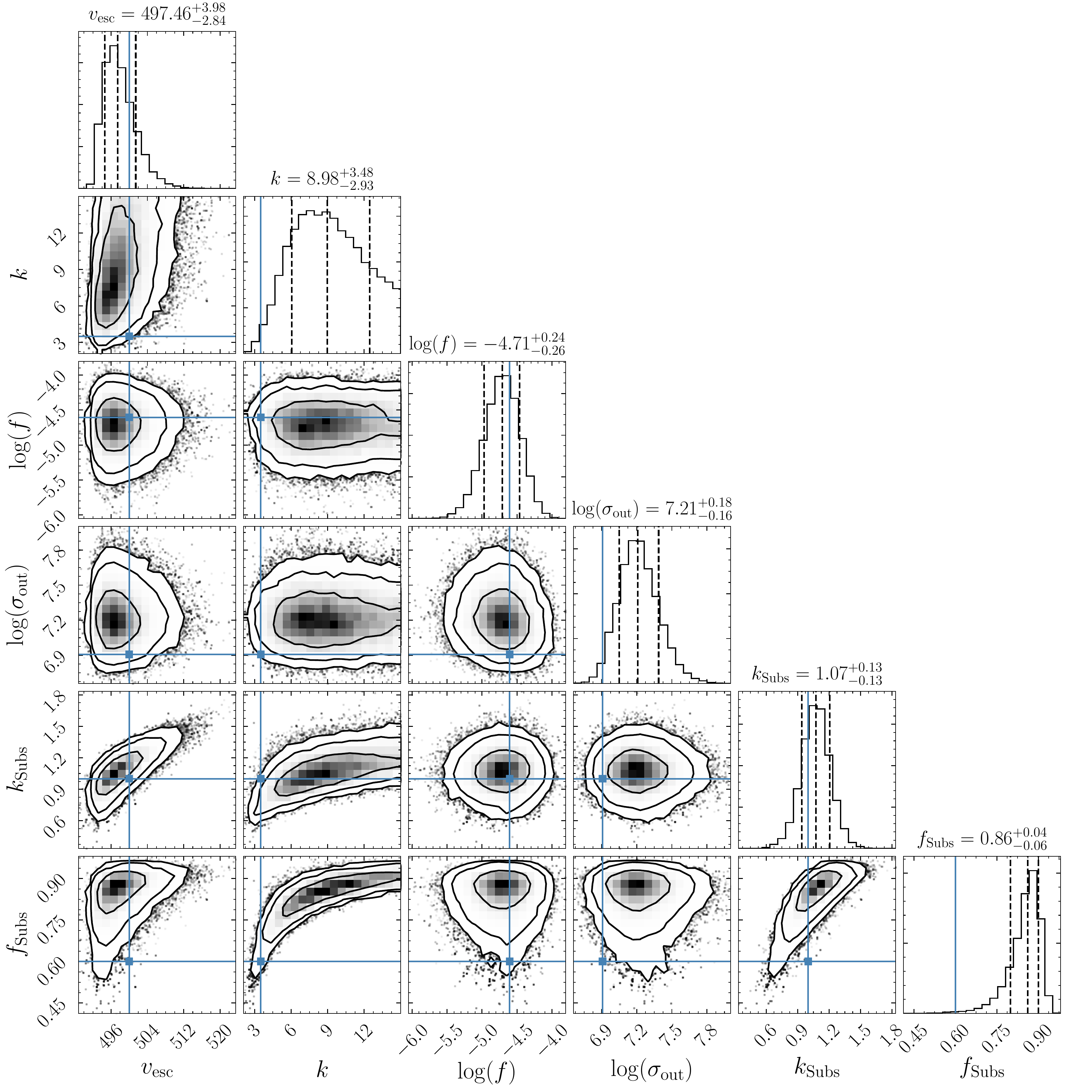} 
   \caption{Corner plot of the run assuming two bound components, $\vmin=$ 300 km/s, and no errors. The 2D contours are the 68\%, 95\%, and 99\% containment regions, and blue lines indicate the true parameter values. the See discussion in \Sec{sec:no_errors}.  }
   \label{fig:corner_plot_no_error}
\end{figure}

\begin{figure}[h] 
   \centering
	\includegraphics[width=0.65\textwidth]{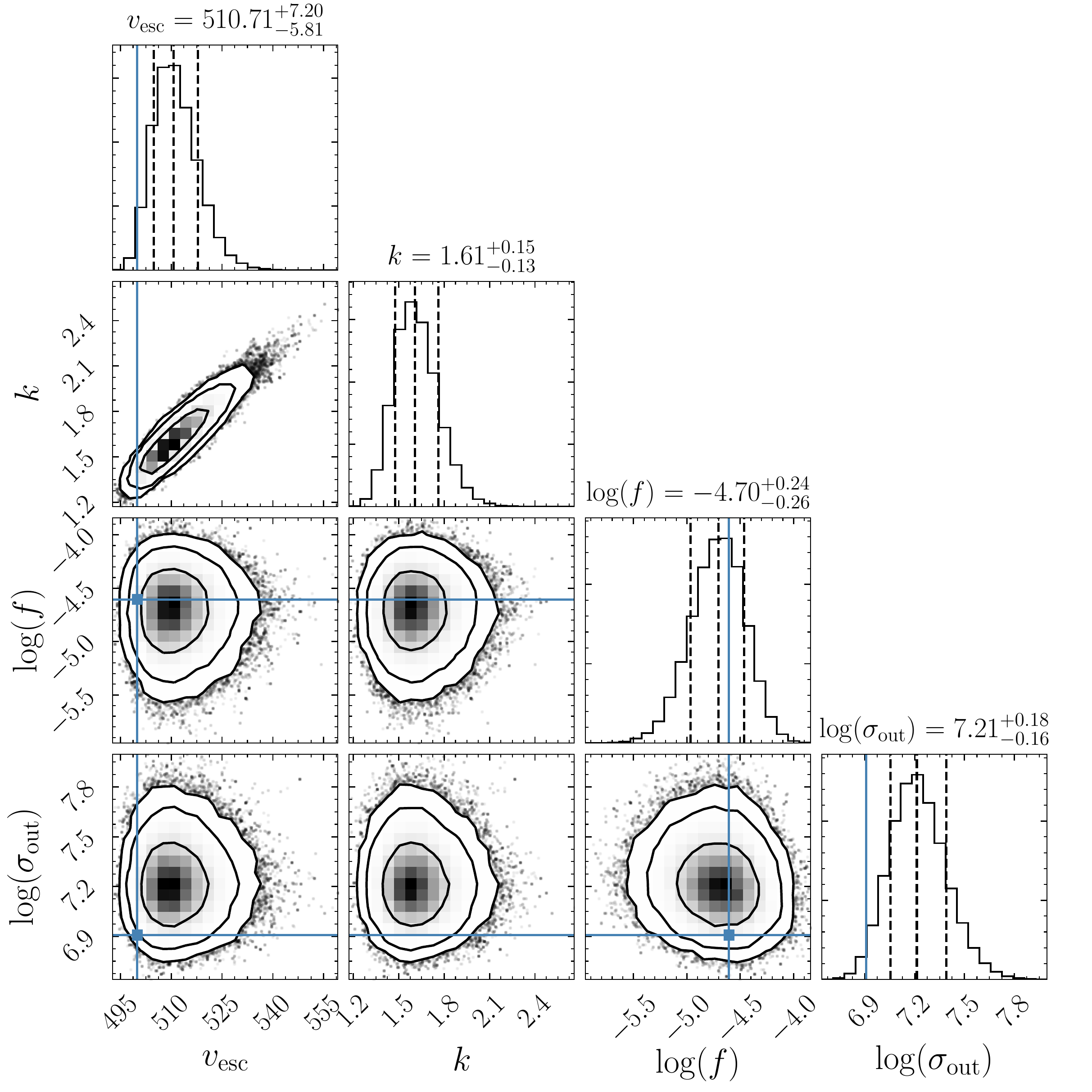} 
   \caption{Corner plot of the run assuming one component, $\vmin=$ 300 km/s, and no errors. The 2D contours are the 68\%, 95\%, and 99\% containment regions. The escape velocity is biased towards larger values compared to the true $\vesc=500$ km/s. See discussion in \Sec{sec:no_errors}.  }
   \label{fig:corner_plot_no_error_1component}
\end{figure}

In \Fig{fig:corner_plot_no_error} and \Fig{fig:corner_plot_no_error_1component}, we show the corner plots for mock data generated with no observational errors. The mock data set is the same fiducial data set discussed in \Sec{sec:no_errors}.

\pagebreak
\newpage

\pagebreak
\newpage

\subsection{5\% errors}

\begin{figure*}[h] 
   \centering
	\includegraphics[width=0.95\textwidth]{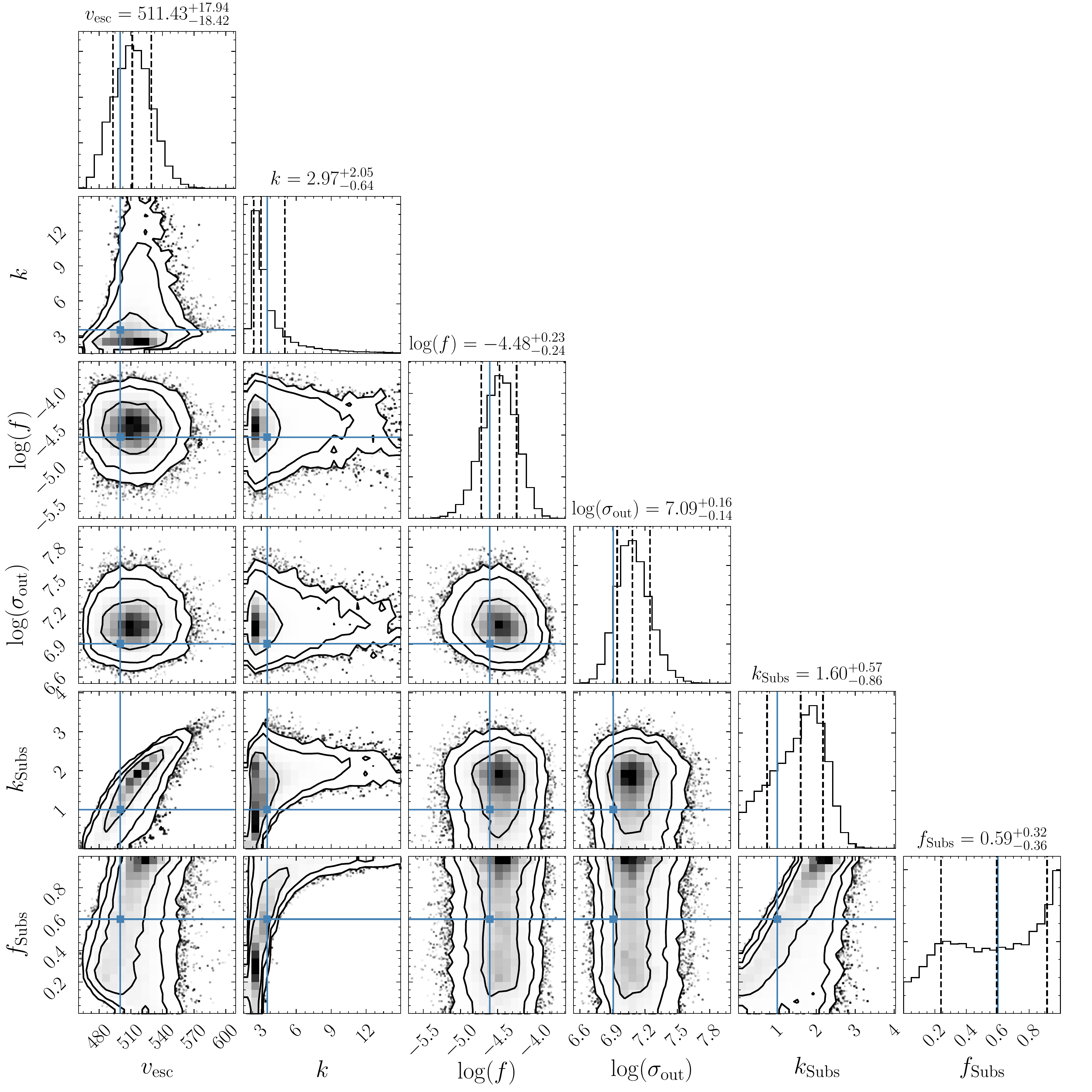} 
   \caption{Corner plot of the run assuming two bound components, $\vmin=$ 300 km/s, and percentage errors of 5\%. See discussion in \Sec{sec:percent_errors}.}
   \label{fig:corner_plot_5_percent}
\end{figure*}

\begin{figure*}[h] 
   \centering
	\includegraphics[width=0.65\textwidth]{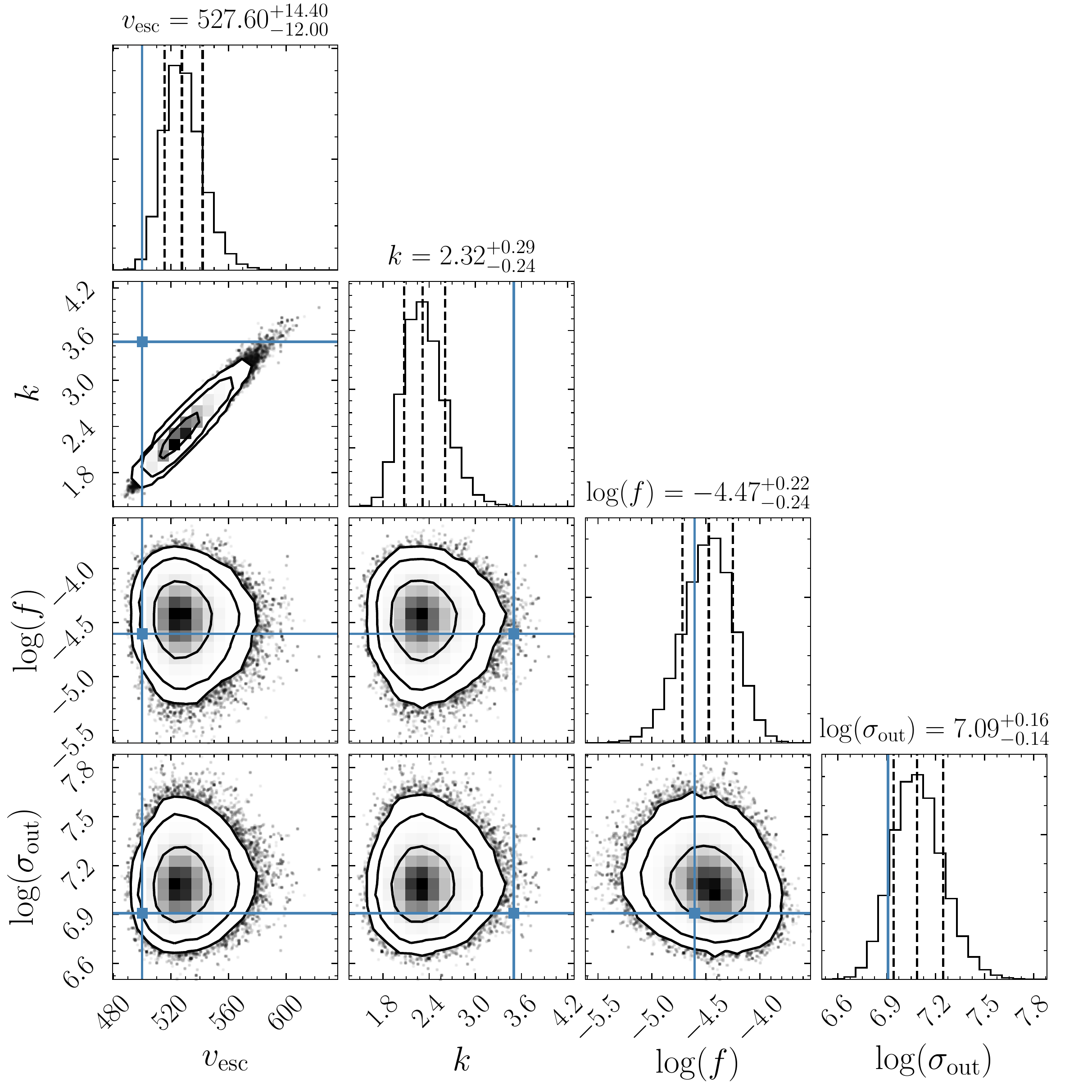} 
   \caption{Corner plot of the run assuming one component, $\vmin=$ 300 km/s, and percentage errors of 5\%. The escape velocity is biased towards larger values compared to the true $\vesc=500$ km/s. The true $k$ shown here is that of the component with $k=3.5$. See discussion in \Sec{sec:percent_errors}.}
   \label{fig:corner_plot_5_percent_1component}
\end{figure*}

In \Fig{fig:corner_plot_5_percent} and \Fig{fig:corner_plot_5_percent_1component}, we show the corner plots for mock data generated with 5\% observational errors. The mock data set is the same fiducial data set discussed in \Sec{sec:percent_errors}.

\pagebreak
\newpage

\pagebreak
\newpage

\subsection{20 km/s errors}

\begin{figure*}[h] 
   \centering
	\includegraphics[width=0.95\textwidth]{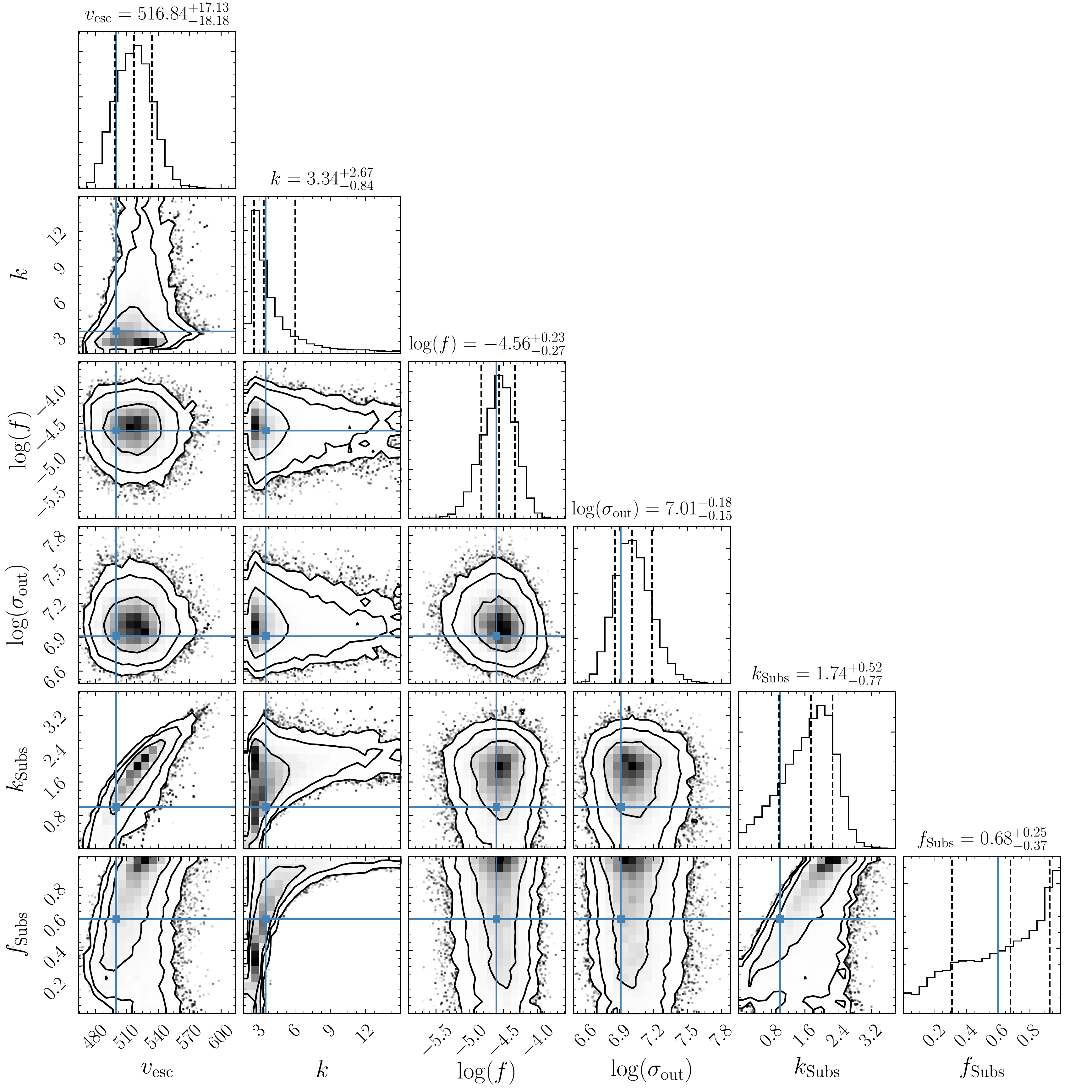} 
   \caption{Corner plot of the run assuming two bound components, $\vmin=$ 300 km/s, and absolute errors of 20 km/s.}
   \label{fig:corner_plot_20_absolute}
\end{figure*}
\pagebreak
\newpage

\begin{figure*}[h] 
   \centering
	\includegraphics[width=0.65\textwidth]{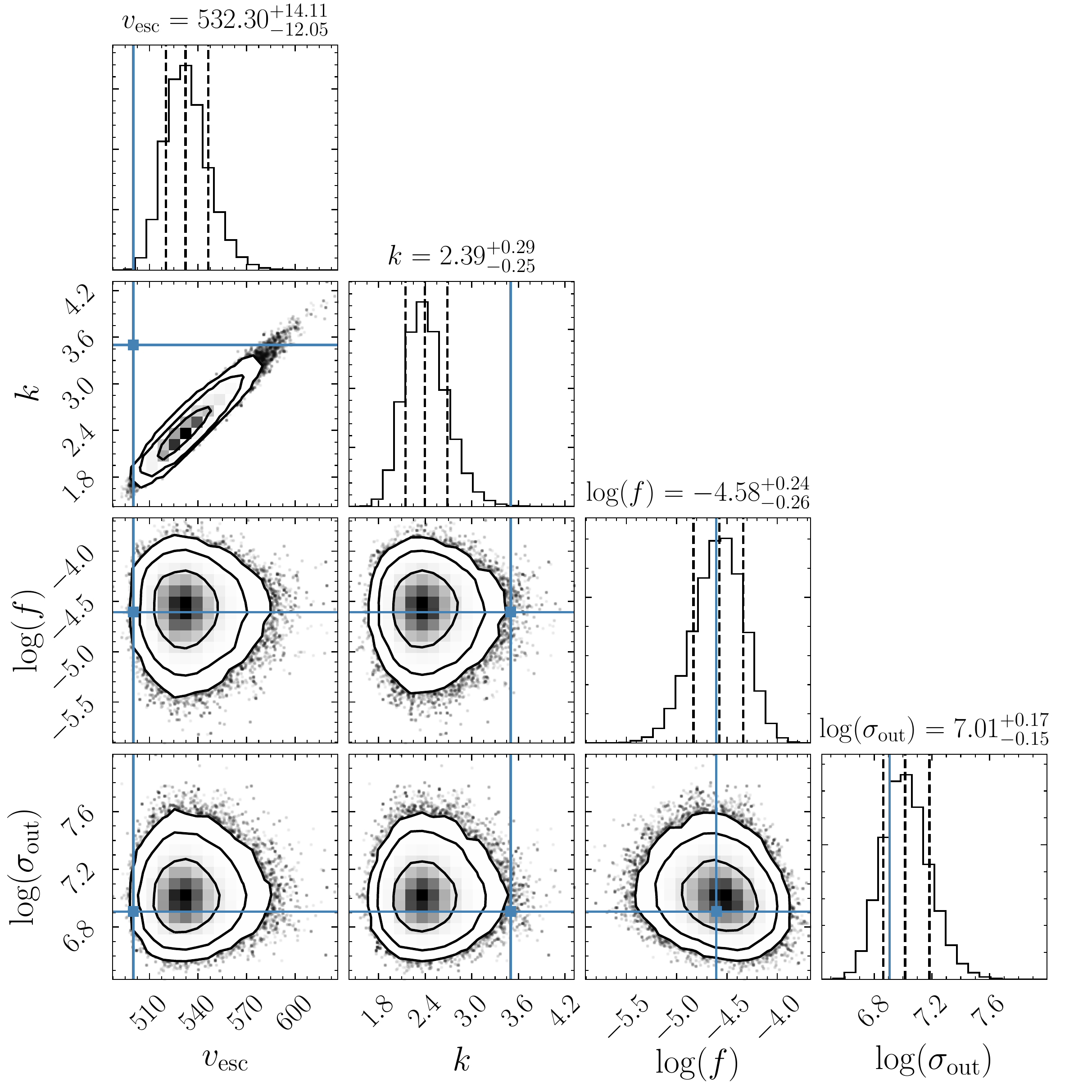} 
   \caption{Corner plot of the run assuming one component, $\vmin=$ 300 km/s, and absolute errors of 20 km/s. The true $k$ shown here is that of the component with $k=3.5$. }
   \label{fig:corner_plot_20_absolute_1component}
\end{figure*}
\pagebreak
\newpage


\section{Further case studies}
\label{sec:app_checks}

In \Sec{sec:other_cases}, we summarized case studies with two scenarios: where the slopes of the components are more similar, and when the measurement errors are larger. We provide the fit results here.


\subsection{Similar slopes}
\label{sec:different_k}

To test the case of similar slopes, we considered mock data with $k_{S}$ = 1 and $k=2$. For $\vmin = 300$ km/s, we find $\vesc = 499^{+13}_{-12}$ km/s from the two-component fit. This is similar to the previous analyses shown in \Fig{fig:mock_sims_percent} for $k_{S} = 1$ and $k=3.5$. When $\vmin = 400$ km/s, we find that including the second component does not reproduce the true values quite as well: there is an underestimation of $\vesc = 488^{+13}_{-8}$ and the posterior for $k_{S} = 0.39^{+0.51}_{-0.28}$ is driven to small values. However, the results are still consistent within the errors.

\begin{figure*}[h] 
   \centering
	\includegraphics[width=0.95\textwidth]{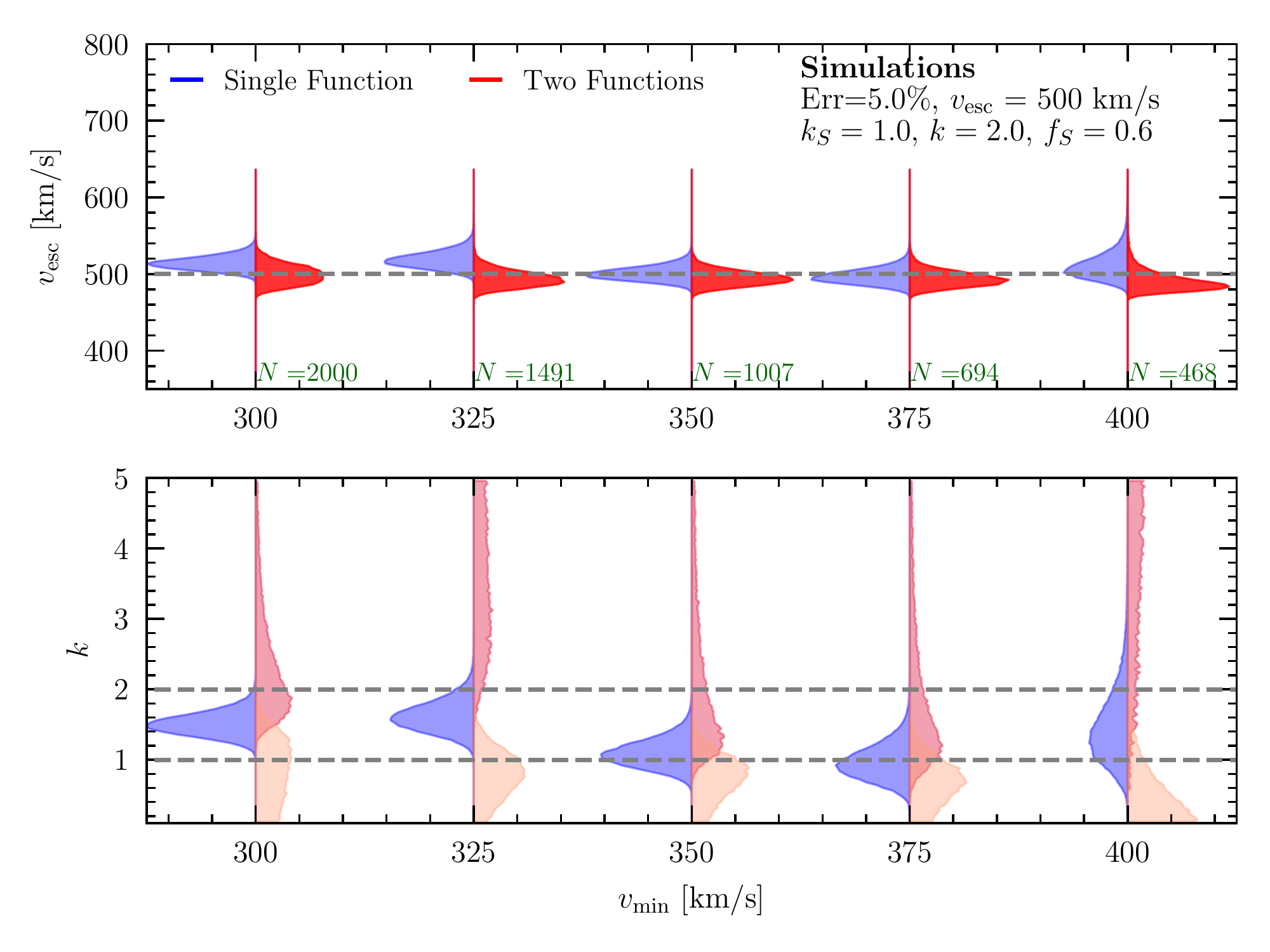} 
   \caption{Similar to \Fig{fig:mock_sims_percent}, but where the slopes of the two substructure components are $k_{S} = 1$ and $k=2$. The stellar speeds are again sampled from a Gaussian distribution with a dispersion of $5\%$ of the true speed. With more similar slopes of the two components, the single-component fit is not quite as discrepant from the true value, and in fact does a better job for large $\vmin$.}
   \label{fig:mock_sims_slopes}
\end{figure*}

\begin{figure*}[t] 
   \centering
	\includegraphics[width=0.47\textwidth]{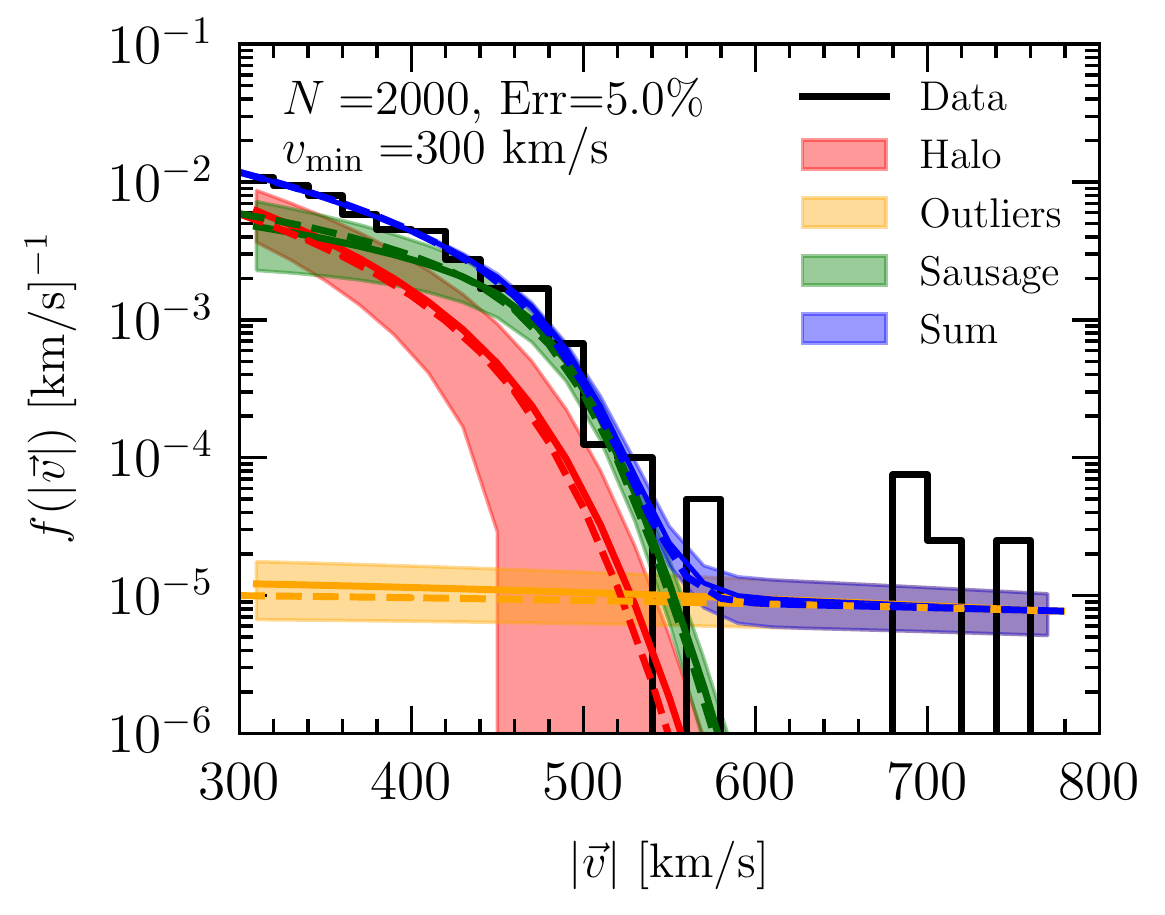} 
	\includegraphics[width=0.47\textwidth]{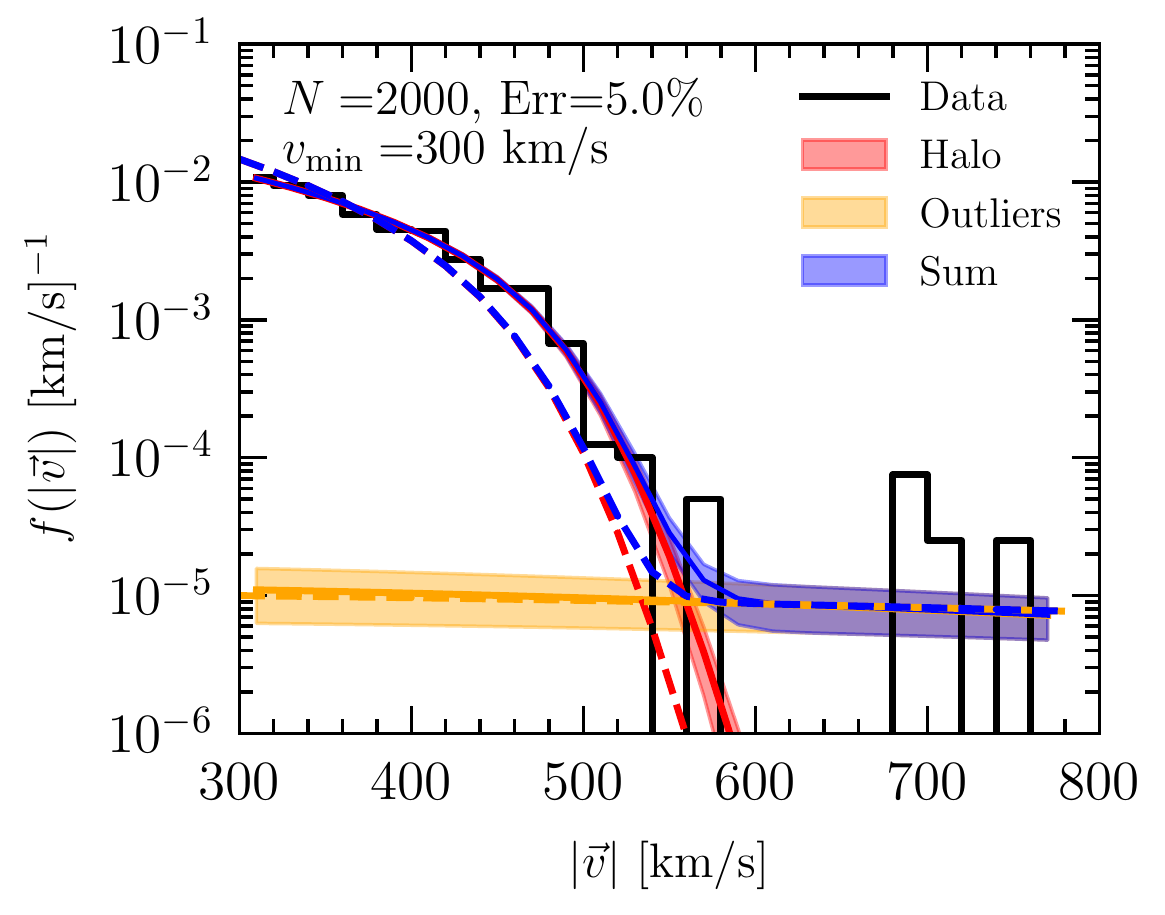} 
   \caption{Similar to \Fig{fig:mock_overfit_no_errors}, but where the slopes of the two substructure components are $k_{S} = 1$ and $k=2$. The stellar speeds are again sampled from a Gaussian distribution with a dispersion of $5\%$ of the true speed. The left panel shows the two-component fit, with the right panel is the single component fit.}
   \label{fig:data_sims_slopes}
\end{figure*}

\begin{figure*}[h] 
   \centering
	\includegraphics[width=0.65\textwidth]{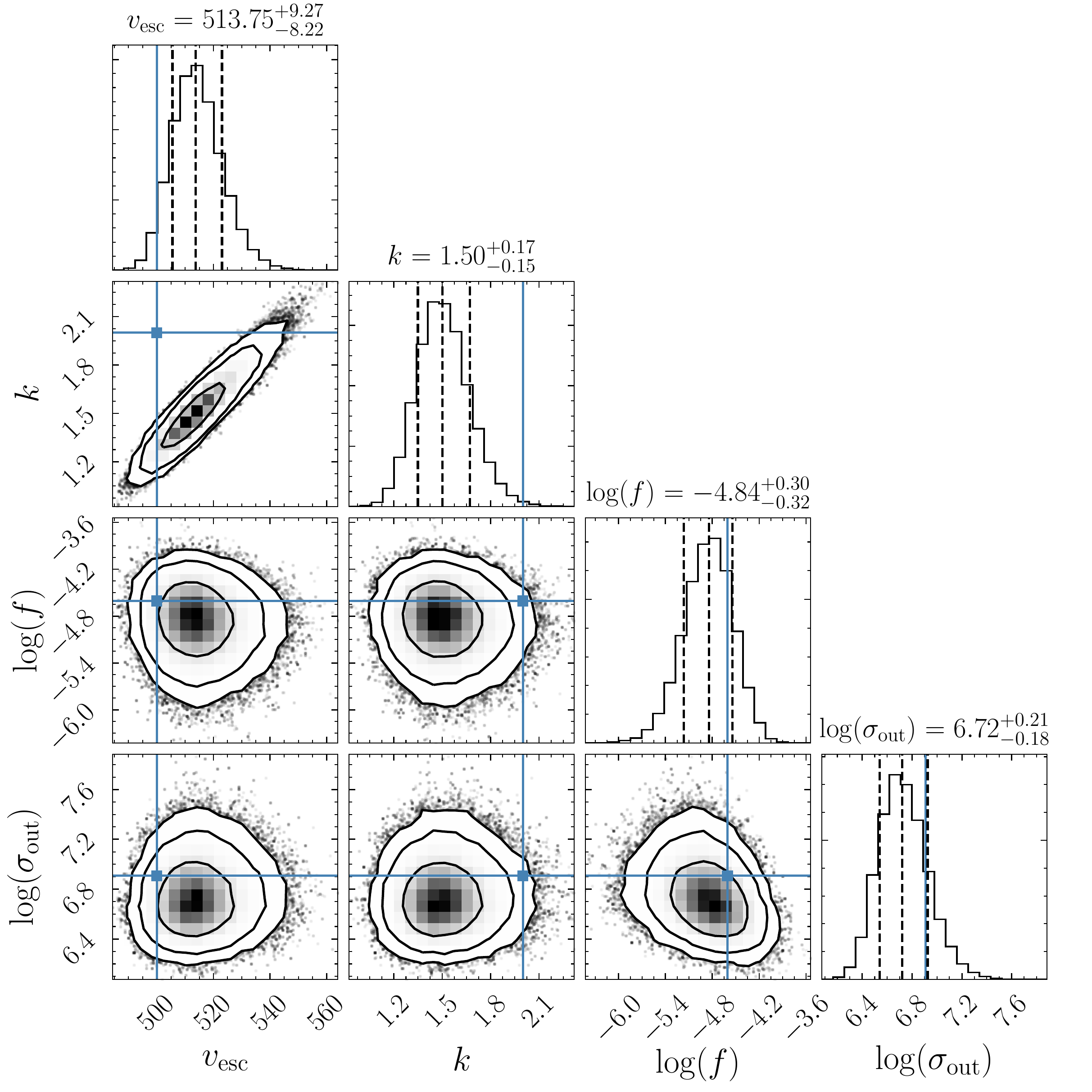} 
   \caption{Corner plot of the run fitting the mock data with one component, with $\vmin=$ 300 km/s. The mock data is generated with two substructure components with $k_{S} = 1$ and $k=2$, and percentage errors of 5\% on the speed. See discussion in \Sec{sec:other_cases}.}
   \label{fig:corner_plot_5_percent_1component_low_ks}
\end{figure*}

\begin{figure*}[h] 
   \centering
	\includegraphics[width=0.95\textwidth]{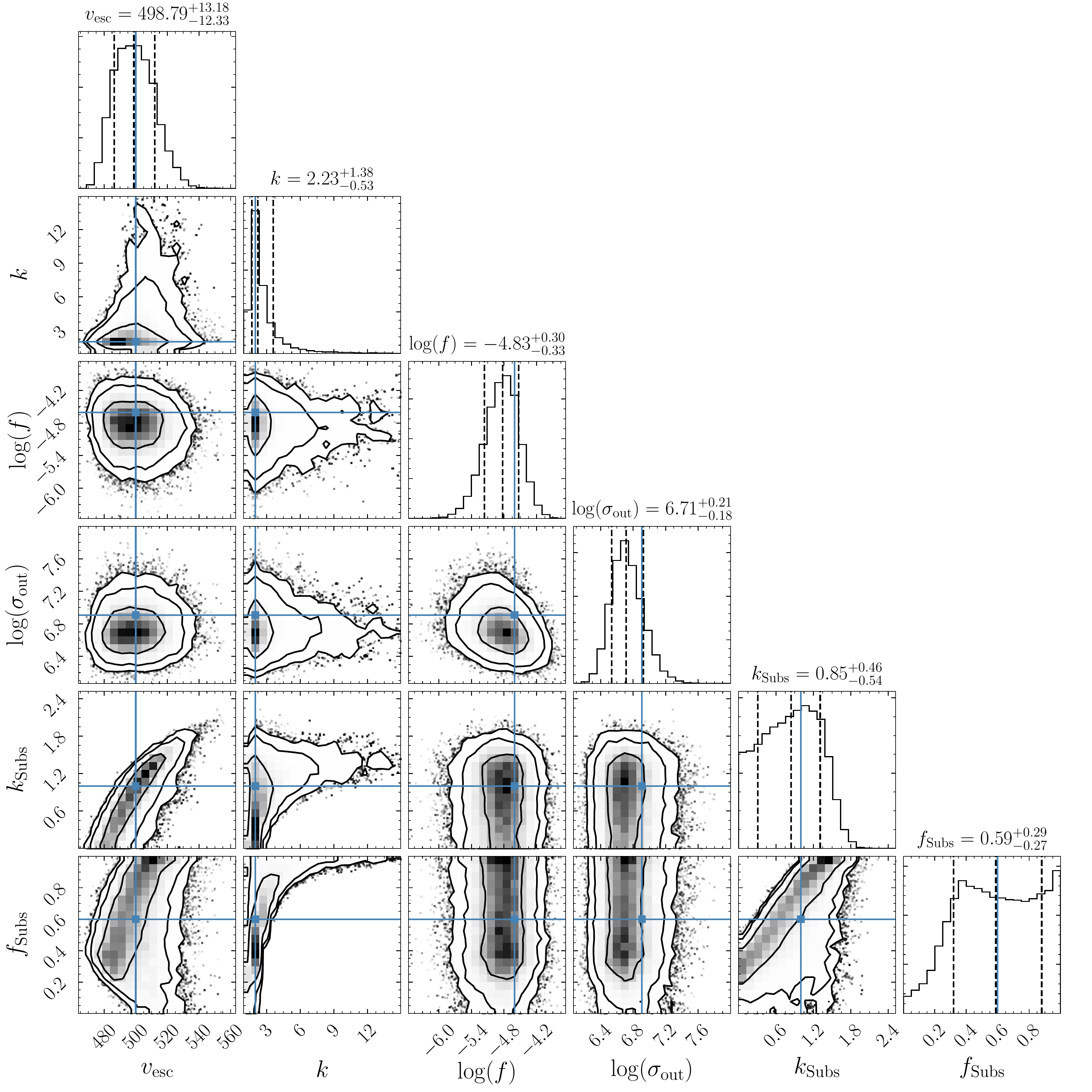} 
   \caption{Corner plot of the run fitting the mock data to two components, with $\vmin=$ 300 km/s. The mock data is generated with two substructure components with $k_{S} = 1$ and $k=2$, and percentage errors of 5\% on the speed. See discussion in \Sec{sec:other_cases}. }
   \label{fig:corner_plot_5_percent_low_ks}
\end{figure*}
\pagebreak
\newpage

\FloatBarrier
\subsection{Impact of large errors \label{app:large_errors}}

\begin{figure*}[ht] 
   \centering
	\includegraphics[width=0.95\textwidth]{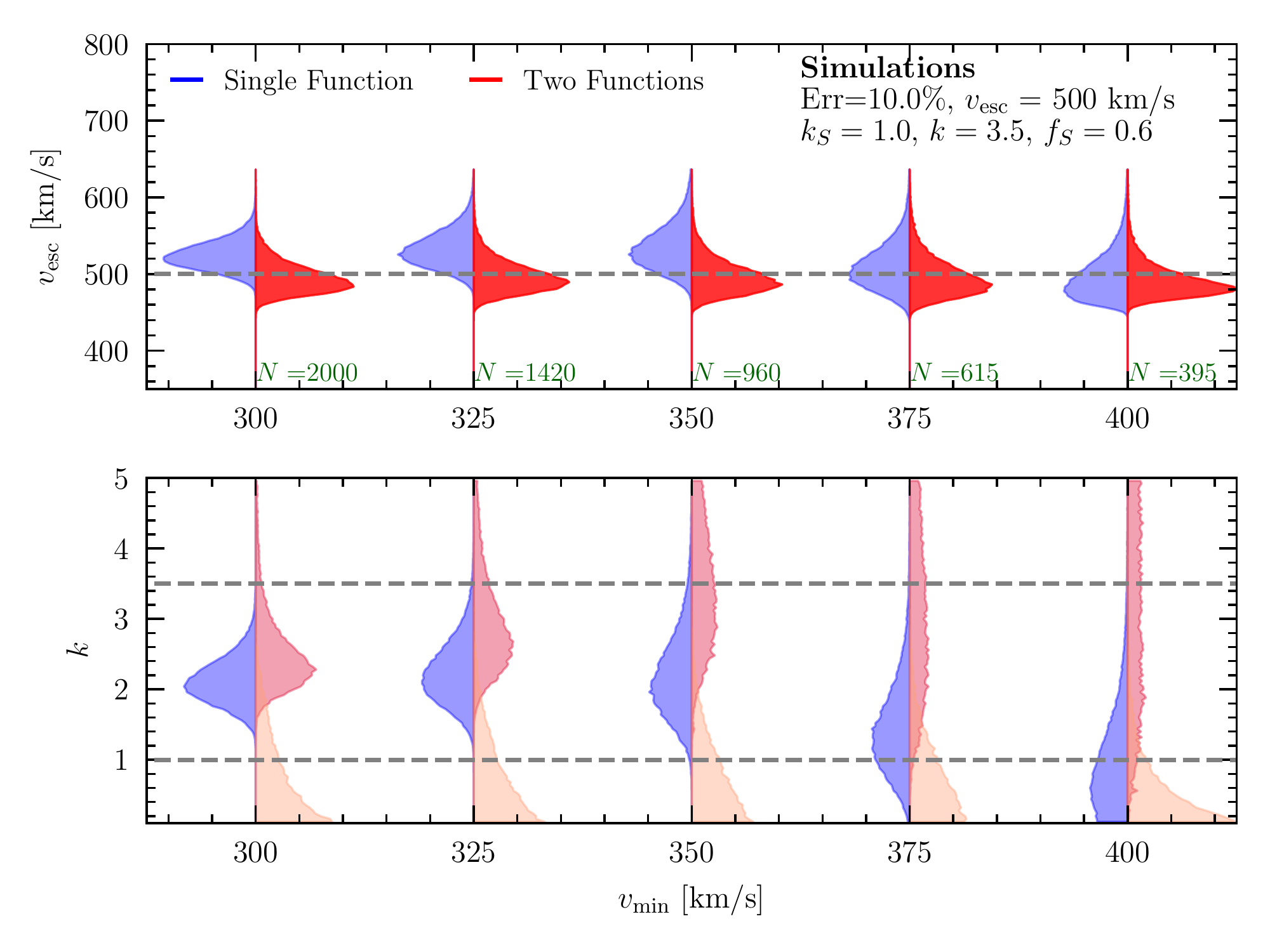} 
   \caption{Similar to \Fig{fig:mock_sims_no_err}, with the stellar speeds now sampled from a Gaussian distribution with a dispersion of $10\%$ of the true speed. \label{fig:vesck_large_err} }
\end{figure*}

With the fiducial model considered in the main text, we also considered a mock data set generated with 10\% errors on the speeds. \Fig{fig:vesck_large_err} shows the fit results for different $\vmin$. The $\vesc$ results are slightly lower than the injected values, but still consistent within one standard deviation. In the two-component fit, $k_{S}$ also tends towards lower values compared to the true values. This may partly be due to the priors selected. When the errors are larger, the behavior of the tail of the distribution is driven more strongly by the errors than by the intrinsic slopes, and it is more difficult to extract $\vesc$. In this scenario, the posterior distribution will be more strongly driven by the priors, with the $1/\vesc$ prior favoring lower $\vesc$ and $k_{S}$.

\begin{figure}[h] 
   \centering
      \includegraphics[trim={0 0 0 0cm},clip,width=0.45\textwidth]{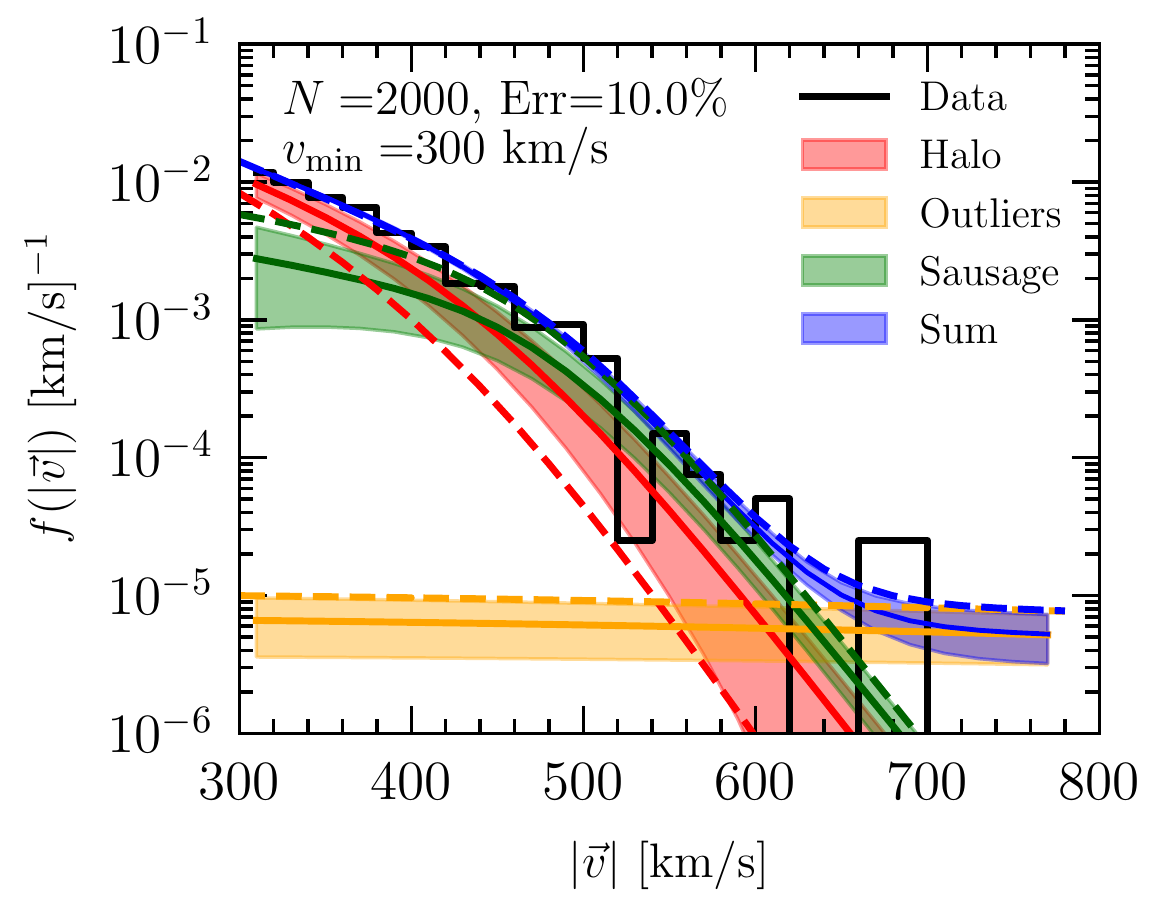} 
      \includegraphics[trim={0 0 0 0cm},clip,width=0.45\textwidth]{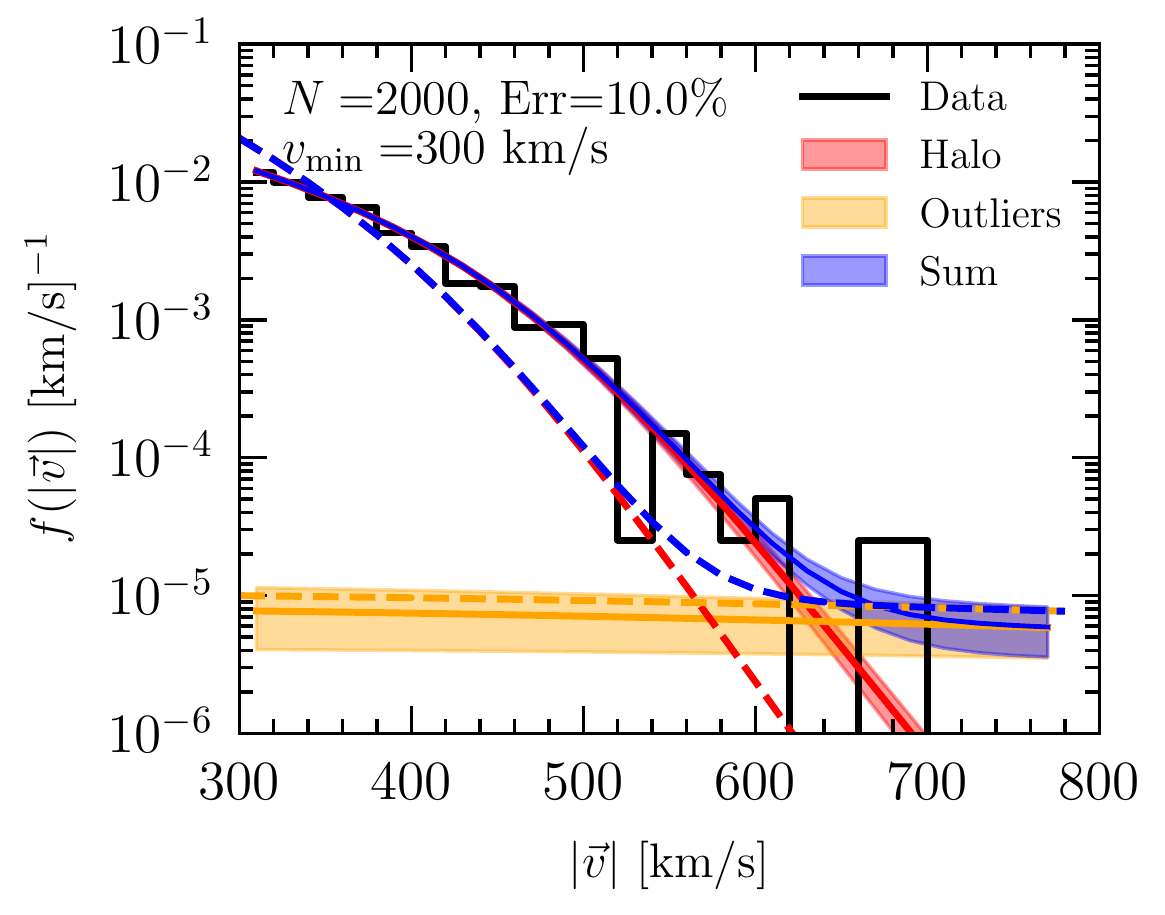} 
   \caption{Similar to the left panel of \Fig{fig:mock_overfit_no_errors}, with the errors generated on the stellar speed sampled from Gaussian distributions with 10$\%$ dispersions.  }
\end{figure}

\end{document}